\documentstyle[11pt,fullpage]{article}
\begin{document}

\def\del{\partial}
\def\wf{W_\infty^f}
\def\be{\begin{equation}}
\def\ee{\end{equation}}
\def\ba{\begin{array}{l}}
\def\ea{\end{array}}
\def\A{{\bar A}}
\def\M{{\overline M}}
\def\eq#1{(\ref{#1})}
\def\D{{\cal D}}
\def\psid{{\psi^\dagger}}
\def\tr{{\rm tr}}
\def\Tr{{\rm Tr}}
\def\aohat{{\widehat{ A_0}}}
\def\l{\lambda}
\def\rhot{{\widetilde \rho}}
\def\U{{\cal U}}
\def\fr{{\rm frac}}

\renewcommand\arraystretch{1.5}

\begin{flushright}
TIFR-TH-94/18
\end{flushright}
\begin{center}
\vspace{3 ex}
{\Large\bf
String Field Theory of Two-Dimensional}\\
\vspace{1 ex}
{\Large\bf QCD on a  Cylinder: A Realization of}\\
\vspace{1 ex}
{\Large\bf $W_\infty$ Current Algebra}\\
\vspace{20 ex}
Avinash Dhar, Porus Lakdawala, Gautam Mandal and Spenta R. Wadia\\
Tata Institute of Fundamental Research \\
Homi Bhabha Road, Bombay 400 005, INDIA \\
\vspace{35 ex}
\bf Abstract\\
\end{center}
\vspace{2 ex}
We consider 2-dimensional QCD on a cylinder, where space is a circle
of length $L$. We formulate the theory in terms of gauge-invariant
gluon operators and multiple-winding meson (open string) operators.
The meson bilocal operators satisfy a $W_\infty$ current algebra. The
gluon sector (closed strings) contains purely quantum mechanical
degrees of freedom. The description of this sector in terms of
non-relativistic fermions leads to a $W_\infty$ algebra. The spectrum
of excitations of the full theory is therefore organized according to
two different algebras: a wedge subalgebra of $W_\infty$ current
algebra in the meson sector and a wedge subalgebra of $W_\infty$
algebra in the glueball sector. In the large $N$ limit the theory
becomes semiclassical and an effective description for the gluon
degrees of freedom can be obtained. We have solved the effective
theory of the gluons in the small $L$ limit. We get a glueball
spectrum which coincides with the `discrete states' of the (Euclidean)
$c=1$ string theory. We remark on the implications of these results
for (a) QCD at finite temperature and (b) string theory.
\thispagestyle{empty}
\newpage
\noindent{\large\bf 0. Introduction}

\vspace{5 ex}

Soluble models of string theory are not abundant and this greatly
limits our ability to tackle many of the important problems in string
theory. In the last several years there has been considerable amount
of work in the $c=1$ string theory \cite{BOOK}. However this theory,
being a closed string theory in two dimensions, has only one
propagating degree of freedom and hence does not provide an
opportunity to understand the physics of the infinite tower of massive
states which are in higher dimensions responsible for some of the most
remarkable properties of string theory. Recently there has been a
trend to explore lower dimensional gauge theories
\cite{WITTGAUGE,GROSSGAUGE,DOUGLAS,MINAHAN}.
One motivation for this is to provide
new models of string theory. In a previous paper \cite{DMWQCD}\ we had
considered the problem of QCD with fermions in a two dimensional
spacetime which is a plane. We formulated the theory in terms of
gauge-invariant open string operators (which physically represent a
meson) satisfying a $W_\infty$ algebra. We reinterpreted the infinite
tower of mesons (a result originally due to 'tHooft \cite{THOOFT2}) as
perturbations around a classical ground state that form a
representation of the wedge subalgebra $W_{\infty +} \otimes W_{\infty
+}$ of the original $W_\infty$ algebra, thus providing an example of
an infinite dimensional algebra that organizes the massive levels of
string theory.

In this paper we discuss the more complicated example of QCD (with
fermions) on a two-dimensional cylinder where space is a circle of
length $L$ and time is non-compact.  Unlike the case of the
two-dimensional plane, QCD$_2$ on a cylinder has {\em dynamical} gauge
fields, besides the Coulomb force between the quarks, giving rise to a
much larger class of gauge-invariant operators. Let us denote by
$\Gamma_{xy}^{(n)}$ the path which starts from point $x$ on the circle
and reaches the point $y$ after winding around the circle $n$ times in
an anticlockwise fashion. The non-abelian phase factor corresponding
to this path is given by the operator $U^{(n)}(x,y,t)|_{ab}\equiv (P
\exp[i g \int_{\Gamma_{xy}^{(n)}} A_1(x',t)dx'])_{ab}$.
The present theory
is described in terms of two basic sets of gauge-invariant operators
constructed out of the above operator: (1) the pure gluon operators
$\U(n,t) \equiv {1\over N} tr U^{(n)}(x,x,t)$ and (2) the
multiple-winding `meson' operators $M^{i\alpha, j\beta}_{xy}(n,t) =
{1\over N} \sum_{a,b} \psi_{i\alpha}^a(x,t) U^{ (n) }_{ab}(x,y,t)
\psi_{j \beta}^b(y,t)$ (here $i,\alpha$ and $a$ respectively denote
flavour, dirac and colur indices). Because of the presence of
non-trivial winding numbers, the `mesons' this time satisfy a
$W_\infty$ {\em current algebra} ($\widetilde {W}^f_\infty$) (Sec. 2)
and obey constraints that depend on the $\U(n,t)$ (Sec. 3).  These
constraints define a coadjoint orbit of $\widetilde{W}^f_\infty$ and
lead to a Kirillov construction of the classical action (Sec. 4).

One of the interesting features in the quantum mechanics of our model
is the nature of the fermi vacuum (Sec. 5). We find that there is no
fixed fermi vacuum labelled by a fixed fermi level since the latter is
neither a gauge-invariant notion nor a conserved quantity due to the
anomaly in the axial current. The right vacuum instead is a linear
combination of fermi vacua labelled by all possible values of the
fermi level. The implication of this on the gluon dynamics (which, in
the gauge $A_1=$ constant and diagonal, is described in terms of $N$
non-relativistic fermionic particles moving on a circle) is that
effectively the fermions live on the real line under the influence of
a harmonic oscillator potential and a mutual interaction (Sec. 6).
Some of these phenomena have earlier been observed by Manton
\cite{MANTON} in the context of the Schwinger model on the cylinder.

The description of the gluon sector in terms of non-relativistic
fermions leads to the interesting phenomenon that there are two
$W_\infty$ algebras in our theory; one is the current algebra
$\widetilde {W}^f_\infty$ of the `mesons' described earlier and the
second is the $W_\infty$ algebra that is formed by fermion bilinears
where the `fermions' now mean those in the gauge sector. Indeed since
these latter fermions live on a circle whose length goes as $1/L$ (as
against the quarks which move in real space of radius $\sim L$) there
is a rather interesting duality between these two sets of fermions. In
the small $L$ limit the `meson' fluctuations (excitation energies
$\sim 1/L$) are effectively frozen (Sec. 8) and the $W_\infty$ of the
gluon sector becomes operative. We find the `glueball spectrum' in
this limit and show that it coincides with the `discrete states' of
the $c=1$ string theory and is arranged by the $W_\infty$ algebra
\cite{WINF}. In the large $L$ limit, on the other hand, it is the
mesons which fluctuate and the gluons are frozen.  In that limit the
physics reduces to that of the planar model where the meson
fluctuations are arranged by a $W_\infty$ algebra \cite{DMWQCD}\ which
is the zero mode part of the current algebra (Sec. 9). For
intermediate values of $L$ we indicate how meson fluctuations around
the ground state form a representation of a wedge subalgebra of
$\widetilde {W}^f_\infty$ (Sec. 7) although the interpretation of them
as spacetime particles and how they interact with the glueballs
remains an open question.

\vspace{5 ex}

\noindent
{\large\bf 1. The action and the hamiltonian}

\vspace{5 ex}

We consider the gauge group $U(N)$. As usual we denote the gauge
fields, which are hermitian matrices, by $A^{ab}_\mu$ where $\mu =
0,1$ is the Lorentz index and $a,b = 1,2, \ldots, N$ are colour
indices. The fermions are denoted by $\psi^a_{i\alpha}$ where $i=1,2,
\ldots,n_f$ is the flavour index and $\alpha= {\pm 1} $
is the dirac index.
We consider one space and one time dimension where the space dimension
is a circle of length $L$.

The theory is described by the following path-integral
\be
\ba
Z = \int \D A_0(x,t) \D A_1(x,t) \D\psi(x,t) \D\psid(x,t)
\exp[iS(A_0,A_1, \psi,\psid)] \\
S = \int_0^T dt \int_0^L dx \big(- {1\over 4} \tr F_{\mu\nu}
F^{\mu\nu} + \bar\psi \gamma^\mu (iD_\mu) \psi - m\bar\psi
\psi ) \\
F_{01} \equiv E = \del_t A_1 - \del_x A_0 + ig [A_0, A_1],\quad
D_\mu = \del_\mu + ig A_\mu
\ea
\label{one}
\ee
which has the following gauge invariance
\be
\ba
\psi(x,t) \to \psi^\Omega(x,t) \equiv \Omega(x,t) \psi(x,t) \\
A_\mu(x,t) \to A^\Omega_\mu (x,t) \equiv {i\over g} \del_\mu \Omega(x,t)
\Omega^\dagger (x,t)  + \Omega A_\mu \Omega^\dagger
\ea
\label{two}
\ee
where  $\Omega(x,t)$  is an arbitrary  U(N) matrix.   We first fix the
gauge $\del_x  A_1 = 0$.  Using the standard Fadeev-Popov procedure we
can reduce $\D    A_1(x,t)$  to $\D\A(t)$  with  appropriate  Jacobian
factors. We get
\be
\ba
Z = \int \D\A(t) \Delta_1 [\A]
\D A_0(x,t) \D\psi(x,t) \D\psid(x,t) \exp[i (S_0 + S_{coul} + S_F)] \\
\Delta_1 [\A] =  {\rm det} ( i\del_x/g + ad\, \A )  \\
S_0 = \int dt {L \over 2} \big[ \tr (\del_t \A + ig [ A_0^{(0)},
\A])^2 \big] \\
S_{coul} = \int dt \int_0^L dx {1\over 2}
\big[ \tr (D_{\A}\aohat)^2 - 2 g\tr (\rho A_0) \big], \quad
\rho_{ab}(x,t) = \psi^{\dagger b} (x,t) \psi^a(x,t)
\\
S_F = \int dt \int_0^L dx \big[\psid i\del_t \psi +
\psid \gamma^5 (i\del_x - g \A) \psi - m\psid \gamma^0
\psi \big]
\ea
\label{three}
\ee
In the above we have denoted the constant mode of the field $A_0(x,t)$ by
$A_0^{(0)}(t)$ and the rest of it by $\aohat(x,t)$. We still have
residual gauge transformations which are purely time-dependent,
$\Omega(t)$, which we now use to further gauge fix $A_0^{(0)}$ to
zero. Going through the Fadeev-Popov procedure once again we get
\be
\ba
Z = \int \D\A(t) \Delta_1[\A] \D\aohat (x,t) \D\psi(x,t) \D\psid(x,t)
\exp [i (S_0 + S_{coul} + S_F ) ] \\
S_0 = \int dt\; {L\over 2} \tr (\del_t \A)^2
\ea
\label{four}
\ee
alongwith the constraint
\be
L \rho_{0,aa} \equiv  Q_a =0
\label{qeq}
\ee
Here $\rho_{0,ab}$ are the generators of colour rotations on fermions
and are zero modes of $\rho_{ab}(x)$ appearing in \eq{three}, thus
\be
\rho_{0,ab} \equiv (1/L) \int_0^L dx \psi^{\dagger b}
(x) \psi^a(x)
\label{rhoeq}
\ee
The Coulomb term $S_{coul}$ and $S_F$ remain the same as in
\eq{three}.  To proceed further we use the following change of
variables:
\be
\A(t) = V(t) \l(t) V^\dagger (t)
\label{five}
\ee
where the colour matrix $\l(t)$ is diagonal.
The measure $\D\A$ changes as
\be
\ba
\D\A(t) = \D \l(t) \D V(t) \prod_t \Delta^2(\l(t)) \\
\D V(t) \equiv \prod_t \prod_{a\neq b} (V^\dagger(t) dV(t))_{ab} \\
\Delta(\lambda(t)) \equiv \prod_{a < b} (\l_a - \l_b)
\ea
\label{six}
\ee
We have used a constraint $(V^\dagger dV)_{aa} =0 $ to make the change
of variable \eq{five}\ well-defined. Under this the action $S_0$
becomes
\be
\ba
S_0 = \int dt {L\over 2} \big[ \tr (\del_t \l)^2 - \sum_{a\neq b}
(\l_a - \l_b)^2 l_{ab}l_{ba} \big] \\
l_{ab} \equiv (V^\dagger (t) i \del_t V(t))_{ab}
\ea
\label{seven}
\ee
In order to compensate for the change of variable \eq{five}\ in $S_F$
and $S_{coul}$ we make corresponding changes of integration variables
$\psi \to V^\dagger \psi, \aohat \to V^\dagger \aohat V$. The
integration measures are invariant under this change. $S_F$ and
$S_{coul}$ now become
\be
\ba
S_F =   \int dt \int_0^L dx  \big[ \psid i\del_t \psi + \tr (\rho_0 l)
+ \psid \gamma^5 (i\del_x - g \lambda) \psi - m\psid \gamma^0 \psi \big]
\\
S_{coul} =  \int dt \int_0^L dx {1\over 2} \big[ \tr (D_\lambda \aohat)^2
- 2g \tr( \widehat{\rho} \aohat ) \big]
\ea
\label{eight}
\ee
Here we have defined $\widehat {\rho}_{ab} \equiv
\rho_{ab}(x) - \rho_{0,ab}$.  The term $\tr (\rho_0 l)$ appears because
of the change of fermion variable $\psi (x,t) \to V^\dagger(t)
\psi(x,t) $. Note that $l_{aa}$ does not appear in it because of eqn.
(5).  That is consistent with the change of variable (7).  The
functional integration over $V(t)$ requires a careful treatment
\cite{BKZ}.  We get
\be
\ba
\int \prod_t \Delta^2 (\l(t)) \D V(t) \exp
\bigg( i \int_0^T dt {L\over 2} \sum_{a\neq b} \big[
(\l_a - \l_b)^2 l_{ab}l_{ba} + 2 l_{ab} \rho_{0, ba} \big] \bigg)
\\
 ~~~~~~~~~ = \Delta (\l(0)) \Delta (\l(T)) \exp (i S_{0,coul})
\\
S_{0, coul} = - \int_0^T dt {L\over 2} \sum_{a\neq b}
\rho_{0, ab} \rho_{0,ba} (\l_a - \l_b)^{-2}
\ea
\label{nine}
\ee
It is easy  to do  the integration over   $\aohat (x,t)$.   This
produces a  Jacobian factor $ 1/\Delta_1 [  \l(t)] $ which cancels the
Jacobian in the measure in \eq{four}. Note that $\Delta_1 [\A(t)] =
\Delta_1 [\l(t)]$. In the action this integration amounts to replacing
in $S_{coul}$ in \eq{eight}\ $\aohat(x,t)$ by $ -g (D_\l)^{-2}
\widehat {\rho} $. In terms of mode expansions $\psi (x,t) = \sum_n
\psi_n(t) \exp (2\pi inx/L)$ this means replacing in $S_{coul}$ the
classical value
\be
A_{0,n}^{ab} =  g { \rho_{n, ab} \over (2\pi n/gL + (\l_a - \l_b))^2 },
\qquad  n\neq 0
\label{ten}
\ee
Combining the resulting $S_{coul}$ with the off-diagonal zero mode
contribution $S_{0, coul}$ of \eq{nine}\ and collecting all the pieces
we get
\be
\ba
Z = \int \D \l(t) \Delta(\l(0)) \Delta (\l(T)) \D \psi(x,t) \D \psid(x,t)
\exp[i (S_0 + S_F + S_{coul})] \\
S_0 = \int_0^T dt\, {L\over 2} \sum_a (\del_t \l_a)^2 \\
S_F = \int_0^T dt\int_0^L dx \big[\psid i\del_t \psi + \psid
\gamma^5 (i\del_x -g \l) \psi - m \psid \gamma^0 \psi \big] \\
S_{coul} = \sum_{a,b} \int_0^T dt \int_0^L dx\int_0^L dy \rho_{ab}(x)
\rho_{ba}(y) K_{ab}(x-y)
\ea
\label{eleven}
\ee
In the above the kernel $K_{ab} (x)$ is
defined by
\be
\ba
K_{ab}(x) \equiv -g^2 L /(8\pi^2) \sum_{n=-\infty}^\infty
\exp(-i2\pi nx/L)[n + (\l_a - \l_b)/\l_0)]^2,
\qquad a\neq b
\\
K_{aa}(x) \equiv - g^2 L/( 8\pi^2) \sum_{n \neq 0}
\exp(-i2\pi nx/L)/n^2
\ea
\label{twelve}
\ee
where
\be
\lambda_0 \equiv \frac{2\pi}{gL}
\label{lambdaperiod}
\ee
Note that the action in \eq{eleven} is invariant under the ``large''
gauge transformations \cite{MANTON,LARGEGAUGE}
\be
\ba
\Omega_{ab} ( \vec m, x) \equiv \exp[ -i gx m_a \lambda_0] \delta_{ab}
\\
\l^\Omega_a = \l_a + m_a \lambda_0
\ea
\label{large}
\ee
The measure is {\em not} invariant under these transformations because
of the presence of the van der Monde factors $\Delta(\l(0))$ and
$\Delta(\l(T))$. The transformations \eq{large}\ were part of the
residual gauge transformations in the gauge $\del_xA_1 = 0$ and were
broken when we fixed the gauge $A_0^{(0)} =0$. This is because the
mode expansion of $A_0$ is not invariant under the large gauge
transformations (16): $A^{ab}_{0,n} \rightarrow
A^{ab}_{0,n+m_a-m_b}$.  Now that $A_0$ has been integrated out
completely, we can restore this invariance by observing that since the
classical action is invariant, the path integral automatically
invariantizes the van der Monde factors (this can be established by an
argument involving a change of integration variable corresponding to
\eq{large}). The result is that $\Delta(\l(0))$ and $\Delta(\l(T))$
are replaced by
their periodic counterparts $\Delta_P(\l(0))$ and $\Delta_P(\l(T))$:
\be
\Delta_P(\l) \equiv  \prod_{a < b} \sin[\pi(\l_a - \l_b)/\l_0]
\label{sineq}
\ee
The fact that this is the right measure can be inferred in two
steps. First, we know this to be the case for pure Yang-Mills theory
in two dimensions \cite{SRW,GROSSWITTEN}.
In the presence of dynamical fermions,
since the fermion measure is separately invariant under large gauge
transformations, and the new terms $S_F$ and $S_{coul}$ in the action
are also invariant, the invariantization of the measure for the gauge
fields must proceed in an identical fashion, giving rise to the same
formula \eq{sineq}.

The $\Delta_P$  factors at $t=0$ and $t=T$  in the measure ensure that
the initial and   final wavefunctions  are completely  antisymmetrized
with  respect  to the $\l_a$'s   (since to  start  with  they must  be
symmetric with  repsect to permutation of  the  $\l_a$'s on account of
Weyl-symmetry).  This gives rise to the well-known fermionic nature of
these eigenvalues.

\vspace{2 ex}

\noindent\underline{The kernels}\hfill\break
The kernels \eq{twelve} can be explicitly evaluated. In the $(x,y)$
region of integration pertinent to \eq{eleven}, the argument of the
kernel $K_{ab}(x) $ lies in the range $[-L, L]$. In this range
\be
\ba
K_{ab}(x) = - (g^2/4) \exp[igx(\l_a - \l_b)] \big[ L /( 2 \sin^2
[ \pi (\l_a - \l_b)/\l_0] ) - ix {\rm cot}[ \pi (\l_a -\l_b ) /\l_0]
- |x| \big], \; a \neq b
\\
K_{aa}(x) = - (g^2/4) \big[ L / 6 - |x| + x^2/ L \big], \quad x \in
[-L,L]
\ea
\label{thirteen}
\ee

As we have observed earlier,
the action in \eq{eleven}\ is invariant under the
large gauge transformations \eq{large}. We would like to rewrite the
theory in a form that makes this symmetry manifest. Let us introduce
the periodic (invariant under \eq{large}) $\delta$-function
\be
\delta_P(\l - \l_a) \equiv \sum_{n=-\infty}^{\infty} \delta(\l -
\l_a + n\l_0)
\label{sixteen}
\ee
and note that
\be
\int_0^{\l_0} d\l \, \delta_P(\l - \l_a) = 1
\label{seventeen}
\ee
Using this and the invariance of $S_{coul}$ under \eq{large}\ we may
write $K_{a\neq b}(x)$ as
\be
K_{a \neq b} (x) = \int_0^{\l_0} d\l\int_0^{\l_0} d\l' K(\l -\l';x)
e^{igx (\lambda_a-\lambda)} e^{igx (\lambda_b -\lambda')}
\delta_P(\l -\l_a) \delta_P(\l' - \l_b)
\label{eighteen}
\ee
where
\be
K(\l; x) \equiv - (g^2/4) e^{igx\lambda} \big[ L / ( 2 \sin^2
[\pi\l/\l_0]) - i x {\rm cot} [\pi \l/\l_0 ] - |x| \big]
\label{nineteen}
\ee
Note that we cannot use the $\delta_P$-functions in the integrand in
\eq{eighteen}\ to set $\lambda_a = \lambda$ and $\lambda_b = \lambda'$
in the exponential factors since these $\delta_P$-functions only
equate the fractional parts (in units of $\l_0$) of respectively
$\l_a$ and $\l_b$ with $\l$ and $\l'$.  We use the uncancelled integer
(in units of $\lambda_0$) parts of $\lambda_a$ and $\lambda_b$ in the
following to write the contribution of \eq{eighteen}\ to
$S_{coul}$ as
\be
\int_0^T dt \int_0^L dx \int_0^L dy \int_0^{\l_0} d\l \int_0^{\l_0} d\l'
K(\l - \l'; x - y) \sum_{a\neq b} \rhot_{ab}(x,t) \rhot_{ba}(y,t)
\delta_P (\l - \l_a) \delta_P(\l' - \l_b)
\label{twenty}
\ee
where $ \rhot_{ab}(x,t) \equiv \chi^{\dagger b}(x,t) \chi^a(x,t) $ and
$\chi^a(x,t)$ is defined as
\be
\chi^a_{i\alpha}(x,t) \equiv \exp[i gx (\l_a - \l)]
\psi^a_{i\alpha}(x,t)
\label{chi}
\ee
Note that the $\chi$'s are invariant under the large gauge
transformations (16).  Now, the $a\neq b$ sum in
\eq{twenty}\ can be thought of as $\sum_{a,b}$ $ - \sum_{a=b}$. The
latter, when inserted in \eq{twenty}, involves the quantity $K(\l;x)$
in the limit $\l \to 0$. In this limit $K(\l ; x)$ behaves as
\be
K(\l; x) |_{\l \to 0}  = - \frac{g^2}{4} \bigg[
 \frac{L/2}{ (\pi \l/\l_0)^2 } + ( \frac{L}{6} - |x| + \frac{x^2}{L} )
\bigg ] + o(\l)
\label{twotwo}
\ee
Because of the zero charge condition, eqn. \eq{qeq}, the divergent
term in \eq{twotwo}\ does not contribute to \eq{twenty}. The rest of
it exactly cancels the $K_{aa}(x)$ contribution to $S_{coul}$. Hence,
we get
\be
S_{coul} = \int_0^T dt \int_0^L dx \int_0^L dy \int_0^{\l_0} d\l
\int_0^{\l_0} d\l' K(\l - \l'; x - y) \sum_{a,b} \rhot_{ab}(x)
\rhot_{ba}(y) \delta_P (\l - \l_a) \delta_P(\l' - \l_b)
\label{twothree}
\ee

\vspace{2 ex}

\noindent\underline{The bilocal `meson' operator}
\hfill\break
Introducing now the manifestly gauge-invariant bilocal field
\be
M^{i\alpha, j\beta}_{xy}(\l, t) \equiv \frac{1}{N} \sum_a
e^{ig(x-y)\l_a} \delta_P(\l -\l_a) \psi^a_{i\alpha}(x,t) \psi^{\dagger
a}_{j\beta} (y,t)
\label{twofour}
\ee
we may rewrite \eq{twothree}\ in the manifestly gauge-invariant form:
\be
S_{coul} =  \frac{g^2 N^2 }{4}
\int_0^T dt \int_0^L dx \int_0^L dy \int_0^{\l_0} d\l
\int_0^{\l_0} d\l' {\widetilde K}(\l - \l'; x - y) \tr_{df} (M_{xy}(\l ;t)
M_{yx}(\l',t) )
\label{twofive}
\ee
where
\be
{\widetilde K} (\l; x) =  \frac{L/2}{ \sin^2 (\pi\l/\l_0)} - ix {\rm cot}
(\pi \l / \l_0) -  |x|
\label{twosix}
\ee
and the trace `$\tr_{df}$' is over  dirac and flavour indices. We notice
that in terms of the manifestly gauge invariant operator $M(\l,t)$, an
internal dimension of length $\l_0$ has emerged.

The physical meaning of the  gauge-invariant bilocal fields introduced
in \eq{twofour}\ is transparent  in the exponential representation  of
the periodic $\delta$-function (Poisson's summation formula):
\be
\delta_P(\l -\l_a) = \frac{1}{\l_0} \sum_{n = -\infty}^{\infty}
 e^{i 2\pi n(\l - \l_a)/\l_0}
\label{twoseven}
\ee
Substituting this in \eq{twofour}\ we see that
\be
M^{i\alpha, j\beta}_{xy}(\l, t) = \frac{1}{\l_0}
\sum_{n=-\infty}^{\infty} e^{-i 2\pi n \l/\l_0} M^{i\alpha,j\beta}_{xy}
(n, t)
\label{twoeight}
\ee
where
\be
M^{i\alpha,j\beta}_{xy}(n,t) \equiv \frac{1}{N} \sum_a e^{ig(x -y + nL)
\l_a(t)} \psi^a_{i\alpha}(x,t) \psi^{\dagger a}_{j\beta} (y,t)
\label{twonine}
\ee
is the bilocal ``meson'' field with $n$ windings of the gauge field
around the spatial direction. We note that \eq{twonine} is indeed the
manifestly gauge invariant operator
\be
M^{i\alpha, j\beta}_{xy} (n,t) = {1 \over N} \sum_a \psi^a_{i\alpha}(x,t)
\left(e^{ig\int_{\Gamma^{(n)}_{xy}}A_i(x',t)dx'}\right)_{ab}
\psi^{\dagger b}_{j\beta} (y,t)
\label{twoten}
\ee
where $\Gamma^{(n)}_{xy}$ is a path connecting $x$ and $y$ with $n$
windings around the circle%
\footnote{In higher dimensions $\Gamma_{xy}$ is not characterized by an
integer but $M_{xy}(\Gamma, t)$ still form a closed algebra (see
next section):
\be
[ M_{xy} (\Gamma, t), M_{x'y'}(\Gamma', t) ] =
\delta_{xy'} M_{y'x}(\Gamma' o \Gamma,t) -
\delta_{x'y} M_{x y'}(\Gamma o \Gamma',t)
\ee
where $\Gamma o \Gamma'$ is a curve defined by $(\Gamma o \Gamma')_{xy'}
=  \Gamma_{xy} \Gamma'_{x'y'} \delta_{yx'}$.}%
. In the gauge
\be
A_1(x,t) = V^\dagger(t)~\l(t) V(t), \quad [\l(t)]_{ab} = \delta_{ab}
\lambda_a(t),
\label{gaugedef}
\ee
it simply evaluates to \eq{twonine}.  We also note that
the periodicity of the $\psi$'s in the spatial direction implies the
constraints
\be
M^{i\alpha, j\beta}_{L0} (n,t) =  M^{i\alpha, j\beta}_{00} (n+1,t),
\qquad
M^{i\alpha, j\beta}_{0L} (n,t) =  M^{i\alpha, j\beta}_{00} (n-1,t)
\label{thirty}
\ee
In terms of the variable $\l$ these read as
\be
M^{i\alpha, j\beta}_{L0} (\l,t) =  e^{i2\pi\l/\l_0}
M^{i\alpha, j\beta}_{00} (\l,t),
\qquad
M^{i\alpha, j\beta}_{0L} (\l,t) =  e^{-i2\pi\l/\l_0}
M^{i\alpha, j\beta}_{00} (\l,t)
\label{threeone}
\ee

\vspace{2 ex}

\noindent\underline{The hamiltonian}\hfill\break
We may now write down the Hamiltonian for this problem
\be
\frac{H}{N} = H_{YM} + H_F,
\label{threetwoa}
\ee
where
\be
H_{YM} = \frac{1}{2NL} \sum_a p^2_a,
\label{threetwob}
\ee
and
\be
\ba
H_F = \int_0^{\l_0} d\l \Tr [ i\gamma^5 \del M(\l) - m\gamma^0 M(\l)] -
\frac{g^2 N}{4} \int_0^L dx \int_0^L dy \int_0^{\l_0} d\l
\int_0^{\l_0} d\l'  {\widetilde K} (\l -\l'; x-y) \\
{}~~~~~~~~~~~~~~~~~~~~~~~~~~~~~~~~~~~~~~~~~~~~~~ \tr_{df} (M_{xy}(\l)
M_{yx}(\l'))
\label{threethree}
\ea
\ee
${\widetilde K}(\l; x) $ is given by \eq{twosix}, $\del_{xy} = \del_x
\delta_P( x - y) $ and $\Tr$ stands for trace over dirac and flavour
indices and integration over spatial coordinates. The periodic spatial
$\delta_P$-function is defined similarly to \eq{sixteen}:
\be
\delta_P (x -y )=  \sum_{n = -\infty}^{\infty} \delta (x - y + nL)
\label{threefour}
\ee
The above hamiltonian comes with the prescription that it acts on
wavefunctions which already include the factors $\Delta_P(\l)$
described in \eq{sineq} and are therefore completely antisymmetrized
in the $\l_a$'s.

\vspace{5 ex}

\noindent {\large\bf 2. Operator algebra of the bilocal field}

\vspace{5 ex}

The bilocal fields \eq{twofour}\ (equivalently \eq{twonine}) satisfy a
closed algebra. This can be easily derived using the fermion
anticommutation relation
\be
\{ \psi^a_{i\alpha}(x,t) , \psi^{\dagger b}_{j \beta} (y, t) \}
= \delta^{ab} \delta_{ij} \delta_{\alpha\beta} \delta_P(x-y)
\label{threefive}
\ee
The gauge invariant operators \eq{twonine}(\eq{twoten}) satisfy the following
current algerba
\be
\ba
\big [ M^{i\alpha,j\beta}_{xy}(n,t),
M^{i'\alpha',j'\beta'}_{x'y'}(n',t) \big] = (1/N) \delta^{ji'}
\delta^{\beta \alpha'} \delta_P(y -x')
M^{i\alpha, j'\beta'}_{xy'} (n + n',t)
\\
 ~~~~~- (1/N) \delta^{ij'} \delta^{\alpha \beta'} \delta_P(x-y')
M^{i'\alpha', j\beta}_{x'y} (n + n',t)
\ea
\label{threesix}
\ee
or equivalently the loop algebra
\be
\ba
\big [ M^{i\alpha,j\beta}_{xy}(\l,t),
M^{i'\alpha',j'\beta'}_{x'y'}(\l',t) \big] = (1/N) \delta^{ji'}
\delta^{\beta \alpha'} \delta_P(y -x')\delta_P(\l - \l')
M^{i\alpha, j'\beta'}_{xy'} (\l,t)
\\
 ~~~~~- (1/N) \delta^{ij'} \delta^{\alpha \beta'} \delta_P(x-y')
\delta_P(\l - \l') M^{i'\alpha', j\beta}_{x'y} (\l,t)
\ea
\label{threeseven}
\ee

For $n=n'=0$, we recognize \eq{threeseven}\ as the infinite
dimensional algebra $W_{\infty}^f \equiv W_\infty \otimes U(n_f)
\otimes U(2)$. \eq{threeseven} (\eq{threesix}) involves an infinite
dimensional generalization of the algebra $W_\infty$, which may
appropriately be called $W_\infty$ current algebra. We shall denote
this in the standard notation by $\widetilde W_\infty$ while
$\widetilde W_\infty^f$ will stand for $\widetilde W_\infty \otimes
U(n_f)$. We have seen in \cite{DMWQCD}\ how a central term arises in
$W^f_\infty$ algebra of the bilocal meson fields (in the case in which
$L \to \infty$) when expanded around a given classical background.  A
similar phenomenon occurs in the present case and fluctuations of
$M(\l)$ around any given classical background satisfy a loop
algebra with a central term.  This central term is essentially the
ground state expectation value of the bilocal field $M(\l)$. This
is discussed in detail in Sec. 7.

\vspace{5 ex}

\noindent {\large\bf 3. Constraints}

\vspace{5 ex}

We now show that the zero charge condition, eqn. \eq{qeq}, implies a
quadratic constraint on $M(\l)$. Using fermion anticommunication
relations we can prove the identity
\begin{eqnarray}
(M^2(\l))^{i\alpha,j\beta}_{xy} &=& (1/N) \sum_a \delta_P(\l -\l_a)
(M(\l))^{i\alpha,j\beta} _{xy} \\
&+& (L/N^2) \sum_{a,b} e^{ig(x - y)\l} \delta_P(\l - \l_a)
\delta_P (\l -\l_b) \chi^a_{i\alpha}(x)\chi^{\dagger b}_{j\beta}(y)
\rho_{0, ab}
\label{threeeight}
\end{eqnarray}
where $\rho_{0,ab}$, defined in eqn. \eq{rhoeq}, is the generator of
colour rotations on fermions and $\chi$'s are the gauge-invariant
fermions defined in \eq{chi}.  Here we have used the convenient matrix
notation for the square of $M(\lambda)$, $\sum_{k,\gamma}
\int^L_0 dz \left(M(\lambda)\right)^{i\alpha, k\gamma}_{xz}
\left(M(\lambda)\right)^{k \gamma, j\beta}_{zy} \equiv
\left(M^2(\lambda)\right)^{i\alpha, j\beta}_{xy}$.
Now, because of the $\delta$-functions in the second term in
\eq{threeeight}, the sum over $a,b$ receives contribution only when
$\l_a = \l_b$. However, since the physical states are antisymmetrized
with respect to $(\l_a, \l_b)$ for every pair $(a,b)$, this can happen
only for $a=b$. Hence, the double sum in
\eq{threeeight}\ collapses to only the $a=b$ terms
on physical states. But this contribution also vanishes on physical
states because of the zero charge condition, \eq{qeq}. We, thus,
arrive at the constraint
\be
M^2(\l) = \big[ \frac{1}{N} \sum_a \delta_P(\l -\l_a(t)) \big] M(\lambda)
\label{threenine}
\ee
Let us introduce the density of eigenvalues
\be
\U(\l, t) \equiv \frac{1}{N} \sum_a \delta_P (\l - \l_a(t))
\label{forty}
\ee
By definition, it satisfies the constraint
\be
\int_0^{\l_0} d\l \; \U(\l,t)   = 1
\label{fourone}
\ee
To see the physical meaning of the
eigenvalue density in the pure Yang-Mills sector,
we write using \eq{twoseven}\
\be
\U(\l, t) = \frac{1}{\l_0} \sum_{n= -\infty}^\infty
e^{-i2\pi n \l/\l_0} \U(n,t)
\label{fourtwo}
\ee
Now $\U(n,t)$ is the gauge-invariant loop variable with $n$
windings in the gauge \eq{gaugedef}:
\be
\U(n,t) = \frac{1}{N} \sum_a e^{i2\pi n \l_a(t)/ \l_0} \equiv {1 \over
N}~tre^{ig\int_{\Gamma^{(n)}} A_idx}
\label{fourthree}
\ee
$\Gamma^{(n)}$ is a curve that winds $n$ times around the circle;  $-n$
corresponds to windings in the opposite direction.
Using \eq{forty}\ in \eq{threenine}\ we finally arrive at the
constraint
\be
M^2(\l) =  \U(\l) M(\l)
\label{fourfour}
\ee
Also, from the definition \eq{twofour}\ of $M(\l)$ and the zero
charge condition, eqn. \eq{qeq}, it follows that
\be
\Tr (M(\l) - \U(\l)) = 0
\label{fourfive}
\ee
\eq{fourfour}\ and \eq{fourfive}\ constitute an infinite set
of constraints which the bilocal variable $M(\l)$ must satisfy. As we
shall see later, these constraints specify a coadjoint orbit of the group
$\widetilde W^f_\infty$ and enable us to construct a classical action  in terms
of the bosonized variable $M(\l)$.

\vspace{5 ex}

\noindent {\large\bf 4. The large $N$ limit --- classical action and
equations of motion}

\vspace{5 ex}
In the rest of this paper, we will consider the present model in the
limit $N \to \infty$.  Since quantum fluctuations are suppressed in
this limit, the theory greatly simplifies.  The large $N$ limit is
taken by holding $\bar g = g\sqrt N$ fixed as $N \rightarrow \infty$.
In this limit, then, $\l_0$ grows as $\sqrt N$.  This is because $\l_0
= 2\pi/gL = \sqrt N\; 2\pi/\bar gL \equiv \sqrt N\; \bar\l_0$, where
now we must hold $\bar\l_0$ fixed as $N \rightarrow \infty$.  This
necessitates scaling of $\l$, $\l_a$ and $p_a$ to respectively
$\l/\sqrt N\equiv\bar \l, \; \l_a/\sqrt N\equiv \bar\l_a$ and
$p_a/\sqrt N \equiv \bar p_a$. Note that in terms of the scaled
variables the value of $\hbar$ is $1/N$. Henceforth, we shall
drop the `bar' on $\bar \l_0, \bar\l$ and $\bar \l_a$.  We shall,
however, continue to use $\bar g = g\sqrt N$ to remind ourselves that
in the following $N \rightarrow \infty$ limit has been taken.

We also define here another variable $\M(\l)$ that we shall
actually need in the following discussion:
\be
\ba
\left(\M(\l)\right)^{i\alpha, j\beta}_{xy} = e^{-i\bar g(x-y)\l}
\left ({\widetilde M(\l)}\right)^{i\alpha, j\beta}_{xy}\\
\;~~~ = (1/N\U(\l, t)) \sum_a \delta_P (\l-\l_a)
\chi^a_{i\alpha} (x,t) \chi^{\dagger a}_{j\beta} (y,t)\\
\;~= (1/N\U(\l, t)) \sum_a \delta_P (\l-\l_a) \exp[i \bar g(x-y)(\l
-\l_a)] \psi^a_{i\alpha} (x,t) \psi^{\dagger a}_{j\beta} (y,t)
\ea
\label{fourten}
\ee
where $ {\widetilde M(\l)} =  M(\l) / \U(\l)$.
By construction, $\M(\l)$ is periodic in $x,y$.  It is, however, not
periodic in $\l$.

In terms of $\M(\l)$ the fermionic part of the hamiltonian, $H_F$,
eqn. \eq{threetwoa}, is given in the limit $N \rightarrow \infty$ by
\be
\ba
H_F = \int^{\l_0}_0 d\l\; \U(\l) \; Tr\left[\gamma^5(i\partial - \bar
g\l)~ \M(\l) - m\gamma^0\M(\l)\right] \\
{}~~~~~~~~~~ - {\bar g^2 \over 4} \int^L_0 dx \int^L_0 dy \int^{\l_0}_0
d\l\; \U(\l) \int^{\l_0}_0 d\l' \; \U(\l') \; \bar K(\l - \l'; x-y)\;
tr_{df} (\M_{xy}(\l) \M_{yx} (\l'))
\label{fourtwelve}
\ea
\ee
where
\be
\bar K(\l; x) = e^{i\bar g x\l} \widetilde K(\l;x),
\label{fourthirteen}
\ee
and $\widetilde K(\l;x)$ has earlier been defined in \eq{twosix}.  Also,
in terms of $\M(\l)$ the constraints \eq{fourfour}\ and
\eq{fourfive}\ read
\be
\M^2(\l) = \M(\l),
\label{fourfourteen}
\ee
\be
Tr (1 - \M(\l)) = 0.
\label{fourfifteen}
\ee

\vspace{2 ex}

\noindent\underline{Classical Action and Equations of Motion}:
\hfill\break
\noindent
In order to find out the classical action of the system, we need to
recognize the important fact that $p_a$ and $\M$ do not commute.
Indeed they satisfy the commutation relation
\be
\left[ p_a(t),\; \M(\l,t)\right] = -{i\over N}~\del_{\l_a} \M(\l, t)
\label{foursixteen}
\ee
 From the definition of $\M(\l,t)$, eqn. \eq{fourten}, we see that the
right hand side of \eq{foursixteen}\ does not vanish.
Had this commutator been zero, it would have been easy to write down
independent symplectic forms for the $(\l_a, p_a)$ system (namely,
simply
$\sum_a d\l_a \wedge dp_a$) and for the $\M$ system
(\cite{COADJOINT}) and the
classical action would involve sum of these two symplectic forms. We
shall see below that the $(\l_a, p_a)$ system carries a representation
of the \underline{Virasoro Group} (Diff($S^1$)) while the $\M$
system, as already noted, forms an orbit of ${\widetilde
W_\infty}$. The existence of the commutator \eq{foursixteen}\ owes to
the fact that the Virasoro group acts on the $\M$ system as well,
thus leading to an intertwining of the two group actions. The
situation here is similar to the one encountered in rigid body
dynamics where the space-fixed angular momenta and the linear momenta
do not commute. To disentangle these two group actions we shall adopt
below a procedure analogous to going to the body-fixed angular
momenta.

\vspace{2 ex}

\noindent\underline{Virasoro Action}:\hfill\break
\noindent
The Virasoro group arises as follows.  Let us consider the space of
the $\l_a$'s. The change from one set of $\l_a$'s to another set,
$\l'_a$'s, can clearly be achieved by a reparametrization $f:R \to R,
f(\l_a) = \l'_a \;\forall \, a$. The easiest way to see this is to
think of a two-dimensional plot in which the $\l_a$'s are marked on
the $x$-axis and the $\l'_a$'s are marked on the $y$-axis. The points
$(x,y) = (\l_a, \l'_a)$ provide $N$ number of points for the plot of
$y=f(x)$. In the limit $N \to \infty$ the map is completely specified.
Let us now consider only those changes of $\l_a$'s in which the
integer parts of $\l_a$ do not change. In other words, if we introduce
the useful notation
\be
x = {\rm int}(x) + \fr(x)
\ee
where ${\rm int}(x)$ is the largest integer less than or equal to $x$
and $\fr(x) = x - {\rm int}(x)$, we are looking for functions $f$
which are such that ${\rm int}(f(\l_a)) = {\rm int}(\l_a) \;\forall \,
a$.  It is clear that these functions are actually maps from $S^1 \to
S^1$ and are of winding number one. These functions form a group
(the Virasoro group) under
standard composition and inversion. The reason for restricting to
these functions is that in the discussion of equations of motion we
will be concerned with infintesimal time-evolutions in which ${\rm
int}(\l_a)$'s do not change.

It is easy to see that the density $\U(\l)$ of eigenvalues
which is a single-valued function on the circle $\l \in [0, \l_0]$
changes under a map $f^{-1}: S^1 \to S^1, \l_a \mapsto \l'_a= f^{-1}
(\l_a)$ as follows
\be
\U_f(\l) =  f'(\l) \U(f(\l))
\label{uchangeone}
\ee
where we have defined
\be
\U_f(\l) \equiv \U(\l | \{ f^{-1}(\l_a)  \} ), \quad
\U(\l) \equiv \U(\l | \{ \l_a  \} )
\label{spinonefinite}
\ee
Thus a change of the eigenvalues $\l_a \to f^{-1}(\l_a)$ is
equivalent to a reparametrization $\l \to f(\l)$. Under these
reparametrizations $\U(\l)$ transforms as a spin-one current. It is
convenient to introduce the quantity $\varphi(\l)$ defined by
\be
\U(\l) = \del_\l \varphi(\l)
\label{defphi}
\ee
$\varphi(\l)$ transforms under the above Virasoro group as a scalar:
\be
\varphi'(f(\l)) =  \varphi(\l)
\label{spinzerofinite}
\ee
We now make the important observation that {\em the entire
$\U(\l)$-configuration space is an orbit of the Virasoro group}.
The reason is that, as we have argued, any distribution $\l_a$ of
eigenvalues can be reached from a fiducial distribution $\l_{a,0}$ by
an appropriate $S^1 \to S^1$ map $g^{-1}$, that is
\be
\l_a =  g^{-1}(\l_{a,0})
\label{fidu}
\ee
provided that the integer parts of these two distributions are the
same. Let the densities corresponding to the fiducial distribution and
the arbitrary distribution be called $\U_0(\l)$ and $\U(\l)$
respectively.  By \eq{uchangeone}\ we have
\be
\U(\l) =  \U_0(\l) dg/d\l
\label{fforu}
\ee
By choosing the fiducial distribution $\l_{a,0}$ to be such that
$\U_0$ is uniform (by the normalization condition it must then be
$1/\l_0$), we can explicitly determine the Virasoro transformation
$g^{-1}$ that takes us from $\l_{a,0}$ to $\l_a$. The result is
\be
dg/d\l = \l_0 \U(\l)
\ee
Using \eq{defphi}\ we find
\be
g(\l)/\l_0 =  \int_0^\l \U(\l) = \varphi(\l)  - \varphi(0)
\label{ffound}
\ee
{\em Thus the orbit of the Virasoro group can be parametrized by the
values of the function $\varphi(\l)$.}

We can repeat the above exercise for $\M$. We get
\be
\M(f(\l)|\{\l_a\}) = \M(\l | \{f^{-1}( \l_a )\})
\label{minvariance}
\ee
where, as in \eq{spinonefinite}, we have displayed the dependence of
$\M(\l)$ on $\{ \l_a \}$ for clarity.
Let us choose $f$ to be precisely the transformation
$g^{-1}$ that takes
us from the fiducial distribution $\l_{a,0}$ to $\l_a$, namely
\eq{ffound}. In that case, we see that
\be
\M(g^{-1}(\l) | \{\l_a \} )
= \M(\l | \{\l_{0,a} \} )
\label{mchangeone}
\ee
In other words, if we define a new variable
\be
{\cal M}(\l) =   \M( g^{-1}(\l))
\label{calmdef}
\ee
then according to \eq{mchangeone}\ ${\cal M}(\l)$ does not depend
on the $\{\l_a \}$. More precisely,
\be
[p_a, {\cal M}(\l,t)] = 0
\label{zerocomm}
\ee
{\em This is the disentanglement we were looking for.}

Because of the zero commutation relation \eq{zerocomm}\ it is now easy
to write down the ``$p\dot q$'' term for the whole system as the sum of
the separate ``$p \dot q$'' terms for the $( \l_a, p_a )$
system and the ${\cal M}$ system. The new field ${\cal M}$
satisfies the same constraints as $\M$
\be
{\cal M}^2(\lambda) = {\cal M}(\lambda)
\label{fourthirtythree}
\ee
\be
Tr(1 - {\cal M}(\lambda)) = 0
\label{fourthirtyfour}
\ee
and the $\widetilde W^f_\infty$ algebra
\be
\ba
\left[{\cal M}^{i\alpha, j\beta}_{xy} (\lambda), \;
{\cal M}^{i'\alpha',
j'\beta'}_{x'y'} (\lambda') \right]\\
\,~~~~~~={\l_0 \over N}~ \delta_P(\l -
\l') \; \left\{\delta^{i'j}\delta^{\beta\alpha'} \delta_P(y-x')
{\cal M}^{i\alpha, j'\beta'}_{xy'} (\lambda) -
\delta^{ij'}\delta^{\alpha\beta'}\delta_P(x-y')
{\cal M}^{i'\alpha',j\beta}_{x'y} (\lambda) \right\}
\ea
\label{fourthirtyfive}
\ee
A kinetic term can be constructed for it using a coadjoint orbit
construction, similar to that used in \cite{DMWQCD,COADJOINT}, for the
present case of $\widetilde W^f_\infty$.  In this way we arrive at the
classical action
\be
\ba
S = N \bigg[i \int_\Sigma dsdt \int_0^{\l_0}
d\lambda {\cal M}(\lambda, t, s)
 \left[\partial_t {\cal M}(\lambda, t, s), \partial_s
{\cal M}(\lambda, t, s)
\right] \\
{}~~~~~~~ + \int dt \{ \sum_a p_a \del_t \l_a
-  \left( \sum_a {1\over 2L} p_a^2 + H_F\right) \} \bigg]
\label{fourthirtysix}
\ea
\ee
where the region $\Sigma$ of $(s,t)$ integration is the lower half plane,
$t \epsilon (-\infty, +\infty)$, $s\epsilon(-\infty, 0]$ and we have
the boundary condition that ${\cal M}(\lambda, t, s=0) = {\cal M}(\lambda,
t)$ and ${\cal M}(\lambda, t, s) \rightarrow t$-independent function of
$\lambda$ as $s \rightarrow -\infty$. In the above equation
$H_F$ is given by \eq{fourtwelve}.

We can readily obtain the equations of motion from \eq{fourthirtysix}\
by the standard variational principle.  The allowed (by the
constraints \eq{fourthirtythree} and \eq{fourthirtyfour}) variations
of ${\cal M}(\lambda, t)$ are the $\widetilde W^f_\infty$ group rotations,
\be
\ba
{\cal M}(\lambda, t) \rightarrow e^{iW(\lambda, t)} {\cal M}(\lambda, t)
e^{-iW(\lambda, t)} \\
{}~~~~~~~ = {\cal M}(\lambda, t) + i \left[W(\lambda, t), {\cal
M}(\lambda, t)\right] + o(W^2)
\label{fourthirtyseven}
\ea
\ee
After obtaining the equations of
motion it is convenient to go back to the original variable $\M
(\l, t)$ and reexpress these equations in terms of it.  After these
manipulations we find that \eq{fourthirtysix}\ leads to the equations
\be
\ba
i\left(\partial_t - \frac{1}{L}\sum_a p_a\del_{\l_a}\right)
\M^{i\alpha, j\beta}_{xy} (\l, t) \\
{}~~~ = \left[-\gamma^5(i\partial - \bar g\l) + m\gamma^0, \M(\l,
t)\right]^{i\alpha, j\beta}_{xy} + {\bar g^2 \over 2} \int^{\l_0}_0
d\l'~\U(\l't) \int^L_0 dz \bigg[\bar K(\l'-\l; x-z) \\
{}~~~~~~~~\left(\bar
M_{xz}(\l',t)\M_{zy}(\l,t)\right)^{i\alpha, j\beta} - \bar
K(\l-\l'; y-z)\left(\M_{xz}(\l, t)
\M_{zy}(\l',t)\right)^{i\alpha, j\beta}\bigg],
\label{fourthirtyeight}
\ea
\ee
\be
\del_t \l_a = {p_a \over L}, \quad \del_t p_a =  -\del_{\l_a} H_F,
\label{lameqone}
\ee
where the kernel $\bar K(\l;x)$ is given by \eq{fourthirteen}\ and
\eq{twosix}.  For completeness we reproduce it below:
\be
\bar K(\l; x) = e^{i\bar g x\l} \left[{L/2 \over \sin^2\left({\pi \l
\over \l_0}\right)} - ix \cot \left({\pi\l \over \l_0}\right) - |x|
\right]
\label{fourfortyone}
\ee
Note that the right hand side of the second equation in
\eq{lameqone}\ is non-zero because of
the implicit dependence of $H_F$ on $\{ \l_a \}$ through $\U(\l)$ and
$\M(\l)$.

\vspace{5 ex}

\noindent {\large\bf 5. The Anomaly and the fermi vacuum}

\vspace{5 ex}
In this section we would like to study the effective theory that
describes the dynamics of the gauge fields obtained by `integrating
out' the fermions. For this purpose, in the large $N$ limit that we
are considering, it is enough to restrict ourselves to the fermi
vacuum.  Actually as we shall see, there is really no single fermi
vacuum in this theory in the conventional sense. One has to sum over
fermi vacua labelled by all possible values of the fermi level. This
is intimately connected with the presence of anomaly in the axial
current in this theory as is explained in detail below.  In the
following we shall set the quark mass to zero and consider the case of
a single flavour only.

Let us recall that the fermionic part of the hamiltonian, given in
\eq{fourtwelve}, has a four-fermi interaction term.
As we shall see later, this term can be neglected in the large $N$
limit for the discussion of the effective gauge theory since the error
thus introduced vanishes in this limit. The quadratic fermion part
(linear in $\M$) of \eq{fourtwelve}\ can be rewritten in the massless
case in terms of fermion modes as
\be
H_{F,0} =  -\sum_a \sum_n Tr (\gamma^5 \psi^a_n \psi^{\dagger a}_n
) (n + g\l_a)\\
\label{hamilone}
\ee
We note that the operators $O^a_n=Tr (\gamma^5 \psi^a_n\psi^{\dagger
a}_n)$ commute with $H_{F,0}$ for all $a$ and $n$. In order to
construct the eigenfunctions of $H$ we can therefore start
by constructing simultaneous eigenfunctions of the $O^a_n$. Since we
are interested in the ground state of $H_{F,0}$ we need to consider
only those eigenfunctions of the $O^a_n$ which {\em minimize}
$H_{F,0}$ for any given background values of the $\l_a$. Now an
eigenfunction of the $O^a_n$ which minimizes $H_{F,0}$ in a given
background of $\l_a$'s is nothing but the filled fermi sea in that
background. We therefore proceed to construct the latter.

\vspace{2 ex}

\noindent\underline{The fermi sea in a fixed Yang-Mills background:}
\hfill\break
\noindent To construct the fermi sea we consider, as usual,
the free dirac equation for quarks of each colour and dirac index
\be
[i\del_t + \gamma^5 (i\del_x - \bar g\l_a) ] \psi^a_\alpha = 0
\label{dirac}
\ee
Choosing $\gamma^0 = \sigma_1$ and $\gamma^1 = -i\sigma_2$, we have
$\gamma^5
\equiv \gamma^0 \gamma^1 = \sigma_3$. In this representation the dirac
equation becomes
\be
[i\del_t  + {\rm sgn}(\alpha) (i\del_x - \bar g\l_a)]\psi^a_\alpha = 0
\ee
where $\alpha = \pm 1$ represent the right-, left-moving fermions
respectively. The above equation is solved by
\be
\psi^a_\alpha (x,t) = \sum_n \exp[-iE^a_{\alpha,n}t +
i\frac{2\pi n}{L} x] \psi^a_{\alpha,n}
\ee
where
\be
E^a_{n\alpha} = \frac{2\pi}{L} {\rm sgn}(\alpha) (n +
\frac{\l_a}{\l_0})
\label{fiveone}
\ee
describes the single-particle energy spectrum. For $\alpha = +1$,
$E^a_{\alpha,n}$ increases with $n$. Thus the fermi
(dirac) sea is constructed by filling the levels from $n= -\infty$
upto some fermi level $n^a_{\alpha=1} \equiv n^a_R$. For the
left-moving fermions, similarly, the fermi sea is constructed
by filling from $n = +\infty$ down to the fermi level $n^a_{\alpha
= -1} \equiv n^a_L$.

\vspace{2 ex}

\noindent\underline{Fermi level transforms under large gauge
trnasformations:}\hfill\break
\noindent We recall that under the large gauge transformations (16)
\be
\psi^a_\alpha(x,t) \rightarrow (\psi^\Omega)^a_\alpha(x,t)
= \exp(-i\frac{2\pi m_a x}{L}) \psi^a_\alpha(x,t)
\ee
which implies
\be
(\psi^\Omega)^a_{\alpha,n} = \psi^a_{\alpha, n + m_a}
\label{modeshift}
\ee
Thus the defining condition for the fermi level $n^a_R$
\be
\langle \psi^{\dagger a}_{+1,n} \psi^a_{+1,n} \rangle
= \theta(-n + n^a_R)
\label{fermisearight}
\ee
leads to the following equation for the gauge-transformed fermions%
\footnote{
In \eq{fermisearight} and \eq{fermisearighta} we have introduced
theta functions with integer arguments, defined such that
$\theta(n) = 1 $ for $n  \ge 0$ and $0$ otherwise. Note that with
this definition we have the identity $\theta(n-1) + \theta(-n)=1$.}
\be
\langle (\psi^\Omega)^{\dagger a}_{+1,n}
(\psi^\Omega)^a_{+1,n} \rangle
= \langle \psi^{\dagger a}_{+1,n+m_a} \psi^a_{+1,n+m_a} \rangle
= \theta(- n - m_a + n^a_R)
\label{fermisearighta}
\ee
 From \eq{fermisearight} and \eq{fermisearighta} we see that the fermi
level $(n^a_R)^\Omega$ for the gauge-transformed fermions is given by
\be
(n^a_R)^\Omega =  n^a_R -  m_a
\label{fermitransformright}
\ee
Similar statements are also valid for the fermi level for left-movers
and one gets
\be
(n^a_L)^\Omega =  n^a_L -  m_a
\label{fermitransformleft}
\ee

\vspace{2 ex}

\noindent\underline{The left- and right-moving fermi levels are
related}\hfill\break
We observe
that the left- and right-moving fermi levels are not independent and
they actually get related by the condition $Q_a = 0$ (see equation
\eq{qeq}). The value of
$Q_a$ in the fermi sea is
\be
Q_a =  \sum_{n = -\infty}^{n^a_R} 1 + \sum_{n = n^a_L}^{\infty} 1
\ee
This is a divergent sum and it needs to be regularized. In the
following we will regulate divergent sums over fermionic modes (of the
colour index $a$) by the gauge-invariant regulator $\exp(-\epsilon | n
+ \l_a/\l_0 |)$ which cuts off contributions from large values of
$|n|$
\footnote{This regulator gives results that are non-analytic
in $\l_a$ for finite $\epsilon$. However, the finite
$\epsilon$-independent pieces of regularized quantities can
be shown to be analytic in $\l_a$.}. The result is
\be
Q_a(\epsilon) =  \frac{2}{\epsilon} + (n^a_R + 1 - n^a_L)
\ee
We will define the regularized charge $Q_a^{reg}$
by  subtracting the divergence, that is
\be
Q_a^{reg} =  n^a_R + 1 - n^a_L
\ee
This vanishes only if
\be
n^a_L =  n^a_R + 1 \;\forall \, a
\label{leftright}
\ee
The subtraction that we have used here and the resulting
expression for the regularized charge
is consistent with gauge invariance. There is a beautiful
argument due to Manton \cite{MANTON}\ which shows that a violation of
\eq{leftright}\ leads to a non-gauge-invariant  answer for the total
momentum which  is formally gauge-invariant on states of
total charge zero. His argument is for the abelian (Schwinger) model
but it can be generalized to our case quite easily.
In the following we shall use the notation
\be
n^a_R = n^a_F, \; n^a_L = n^a_F + 1
\ee

\vspace{2 ex}

\noindent\underline{Expectation value of the quark bilinear
in the fermi sea}\hfill\break
Let us  define the notation $|\{n^a_F\}\rangle $ for the fermi sea
in a given background of $\l_a$. The state $|\{n^a_F\}\rangle $ is
defined by the following expectation values
\be
\langle \psi^a_{+1,n} \psi^{a\dagger}_{+1,n} \rangle =
\theta (n - n_F^a -1)
\label{fivethree}
\ee
and
\be
\langle \psi^a_{-1,n} \psi^{a\dagger}_{-1,n} \rangle =
\theta (-n + n_F^a)
\label{fivethreea}
\ee
Using the above results it is easy to calculate the
quantity $\M_{xy}(\l)$ (eqn. \eq{fourten})
in the fermi vacuum $| \{n^a_F \} \rangle$.
We get
\be
\M_{xy}(\l, t) = \frac{1}{L} \sum_{n=-\infty}^\infty e^{i
\frac{ 2\pi n (x-y) }{L}} \M_n(\l, t)
\label{modedef}
\ee
where
\be
\M_n(\l) = \frac{1}{N\U(\l)} \sum_a \sum_m \delta(\l - \xi_a+ m\l_0)
\big[ (\frac{1+ \gamma^5}{2}) \theta(n - m -1)
+ (\frac{1- \gamma^5}{2})
\theta(-n + m)
\big]
\label{mvev}
\ee
In the above equation we have defined a new gauge-invariant quantity
\be
\xi_a = \l_a + n^a_F
\ee
Note that because of non-periodicity of $\M_{xy}(\l, t)$ in $\l$ ($
\M_{xy} (\l + \l_0, t) = e^{i\frac{ 2\pi (x-y)}{L}} \M_{xy}(\l)$), the
modes $\M_n(\l)$ satisfy the following quasi-periodicity condition
\be
\M_n(\l + \l_0) = \M_{n+1} (\l)
\label{quasione}
\ee
The above condition is clearly satisfied by $\M_n(\l)$ calculated
in \eq{mvev}.

It is easy to show that $\M_n(\l,t)$
obtained from \eq{mvev}\
by replacing $\xi_a$ by $\xi_a(t)$ and $\U(\l)$ by $\U(\l,t)$
satisfies the
constraints \eq{fourfourteen} and the \eq{fourfifteen}, and the
equation of motion \eq{fourthirtyeight}.  The quadratic constraint is
satisfied quite trivially. The second (trace) constraint is satisfied
because we have already ensured $Q_a=0$. To see that the equation of
motion \eq{fourthirtyeight} is also satisfied, we note that
$\M_{xy}(\l, t)$ in \eq{modedef}\ is a function of $(x-y)$ only. This
fact, together with the dirac structure of $\M_n(\l,t)$ displayed in
\eq{mvev}, ensures that for massless quarks the right hand side of
\eq{fourthirtyeight} vanishes.  The fact that
the non-linear term in $\M$ does not contribute at all to the equation
of motion is the classical counterpart of the statement that the fermi
sea constructed by ignoring the quartic fermion term is correct in the
$N \to \infty$ limit. With the right hand side equal to zero, equation
\eq{fourthirtyeight} reduces to
\be
(\del_t - {1\over L}\sum_a p_a \del_{\l_a}) \M_n(\l, t) =0
\label{meqtwo}
\ee
It can be easily checked that $\M_n(\l, t)$
obtained from \eq{mvev}\
by replacing $\xi_a$ by $\xi_a(t)$ and $\U(\l)$ by $\U(\l,t)$
satisfies \eq{meqtwo}.

\vspace{2 ex}

\underline{Effective hamiltonian for the gauge fields:}

The effective hamiltonian for the gauge fields is obtained from the
full hamiltonian $H$ by substituting for $\M(\l)$ its value \eq{mvev}
in the fermi vacuum (note that the contribution of the excited states
is subleading in $1/N$). The pure Yang-Mills part of the hamiltonian
of course remains the same. $H_F$ becomes (with quark mass set equal
to zero)
\be
H_F = {\bar g}\l_0 \int_0^{\l_0}d\l\, \U(\l) V(\l)
-\frac{1}{2L\l_0^2}\int_0^{\l_0} d\l\, \U(\l) \int_0^{\l_0} d\l'\,
\U(\l') K(\l, \l')
\label{hf}
\ee
where
\be
V(\l) =  - \sum_{n=-\infty}^\infty (\l
+ n\l_0) {\rm tr}_{df} (\gamma^5 {\M}_n(\l))
\label{potential}
\ee
and
\be
K(\l, \l') = \sum_{n,n'=-\infty}^\infty \frac{ {\rm tr}_{df}
\left[{\M}_n(\l)
{\M}_{n'}(\l') \right]}{ [(\l + n\l_0) -  (\l' + n'\l_0) ]^2}
\label{kernel}
\ee
with $\M(\l)$ given by  \eq{mvev}. Using the
exponential regularization introduced above we get
\be
V_\epsilon(\l) = - \frac{2} {\epsilon} - \frac{1}{12} + V_{reg}(\l)
+ o(\epsilon)
\label{potentialone}
\ee
and
\be
K_\epsilon(\l, \l') = \frac{\pi^2} {\epsilon \sin^2 \frac{\pi}{\l_0}
(\l -\l') } + 2 (\ln \epsilon + C -1) + K_{reg} (\l, \l') +
o(\epsilon \ln \epsilon)
\label{kernelone}
\ee
where
\be
V_{reg}(\l) = \frac{1}{N\U(\l)}
\sum_a \delta_P(\l - \xi_a) (\xi_a/\l_0 + 1/2)^2
\label{potentialtwo}
\ee
\be
K_{reg}(\l, \l') = \frac{1}{N^2 \U(\l)\U(\l')}\sum_{a,b}
\delta_P(\l - \xi_a) \delta_P (\l - \xi_b) \left[ \psi(w_{ab})
+ \psi(-w_{ab}) + w_{ab} \{ \psi'(w_{ab}) - \psi'(- w_{ab}) \}
\right]
\label{kernellast}
\ee
Here $w_{ab} \equiv (\xi_a - \xi_b)/\l_0$ and $C$ is Euler's constant.
Also, $\psi(x) = \frac{d}{dx} \ln \Gamma(x) $ and $\psi'(x) =
\frac{d}{dx} \psi(x)$, $\Gamma(x)$ being the standard gamma-function.
Once again, $\xi_a = \l_a + n^a_F$, the gauge-invariant combination.

Now notice that the above expressions depend on the fermi level
$n^a_F$. The dependence of quantities such as total energy on the
fermi level is of course quite natural. However, since in our problem
the $n^a_F$'s are not gauge-invariant, the question is how one can
possibly have a fermi vacuum specified by a fixed set of $n^a_F$'s and
still not violate gauge-invariance.  The answer, as we shall see
below, is that there is no fixed fermi vacuum in our
theory with a given set of $n^a_F$'s and the total wavefunction for
the full theory (for example for the full ground state) is given by a
sum $\sum_{\{n^a_F\}} | n^a_F \rangle
\otimes \phi_{n^a_F} (\{ \l_a\})$ where $\phi_{n^a_F} (\l_a)$ is the
(lowest energy) eigenfunction of the effective hamiltonian for gauge
fields for given $n^a_F$'s.

We start by observing that $| \{ n^a_F \} \rangle $ is a wavefunction
of the fermions.  To construct wavefunctions of the full theory we
need to consider tensor products of this with wavefunctions of the
gauge fields. Denoting the latter by $\phi_{\{ n^a_F\} } (\{ \l_a \}
)$ we may write for the full wavefunction
\be
\Psi_{n_F} (\psi; \l) \equiv  | \{ n^a_F \} \rangle \otimes
\phi_{\{ n^a_F\} } (\{ \l_a \} )
\label{fullstate}
\ee
Since $| n^a_F \rangle $ is a simultaneous eigenstate of the fermionic
operators $ O^a_n \equiv Tr (\gamma^5 \psi^a_n \psi^{\dagger a}_n )$
and since the $O^a_n$'s commute with the full hamiltonian $H$,
\footnote{These operators commute with the quadratic part of $H_F$
for massless quarks. It can be shown that their commutator with the
quartic fermion term is down by a factor of $1/N$.} the above
wavefunction would also be an eigenstate of $H$ (albeit a
non-gauge-invariant one), provided $\phi_{\{ n^a_F \}}$ satisfies the
equation
\be
\left [\sum_a -\frac{\del_{\l_a}^2}{2N^2 L} +
H_F^{reg} (\{ \xi_a \}) \right ] \phi_{n^a_F}(\{ \l_a \}) =
E \phi_{n^a_F}( \{\l_a \} ), \; \xi_a = \l_a + n^a_F \l_0
\label{phieqone}
\ee
where $H_F^{reg}$ is obtained from $H_F$ in \eq{hf} by substituting
$V_{reg}(\l)$ and $K_{reg}(\l, \l')$ repsectively for $V(\l)$ and
$K(\l, \l')$ in it.

In the above equation the values of $\l_a$ are in the $N$-torus ($
\l_a \in [0, \l_0) $). Since this is a compact region we must
supplement the differential equation \eq{phieqone}\ by boundary
conditions specifying what happens when any one of the eigenvalues
(say $\l_b$) reaches the value $\l_0$. We will now argue that the
appropriate boundary conditions are that both $\phi_{\{n^a_F\}}$ and
its gradient $\del_{\l_c} \phi_{\{ n^a_F \}}$ are continuous:
\be
\ba
\phi_{\{n^a_F \}}( \l_b = \l_0) = \phi_{\{n^a_F +\delta^{ab}\}}
( \l_b =0) \\
\del_{\l_c}\phi_{\{n^a_F \}}( \l_b = \l_0) =
\del_{\l_c}\phi_{\{n^a_F +\delta^{ab}\}}
( \l_b =0), \quad c=1,2,\ldots,b,\ldots,N
\ea
\label{phibc}
\ee
Equation
\eq{phibc}\ is basically a consequence of invariance of the theory
under topologically non-trivial gauge transformations and is connected
with the phenomenon of `spectral flow' \cite{MANTON}. The point is
best illustrated by considering the spectrum, eqn. \eq{fiveone}, of
the dirac operator for each colour $a$, which we reproduce here for
convenience
\be
E^a_{n\alpha} = \frac{2\pi}{L} {\rm sgn}(\alpha) (n +
\frac{\l_a}{\l_0})
\label{spectrumone}
\ee
Consider a one-parameter deformation
\be
\mu \mapsto \l_b (\mu) = \mu \l_0, \quad \mu \in [0,1]
\label{homotopy}
\ee
such that
\be
\l_b(0) = 0, \; \l_b(1) =  \l_0
\label{homtwo}
\ee
Under this flow a right-handed (left-handed) energy level continously
increases (decreases) as $\mu$ increases from 0 towards 1. As $\mu$
reaches 1, however, the spectrum becomes identical to that at $\mu=0$
(namely, $(2\pi/L) \times$ integers). This is in accord with the fact
that $\l_b=0$ and $\l_b=\l_0$ represent the same physical point on the
$\l_b$-circle. We may wish to ask if the
physics is always invariant under the change $(\l_b=0) \to
(\l_b=\l_0)$.  In the pure Yang-Mills theory this is in fact what
happens since $\l_b = \l_0$ can be gauge-transformed to $\l_b =0$. In
the presence of dynamical fermions, however, such a gauge
transformation also changes the fermi momenta (see \eq{modeshift}) and
in particular the fermi levels
(\eq{fermitransformright},\eq{fermitransformleft}).  In this case the
statement of invariance translates to the fact that physics should be
invariant with respect to the change $(\l_b=0, n^b_F) \to (\l_b= \l_0,
n^b_F -1)$.  Indeed every term in $H_F$ remains invariant under this
change because in each of them $\l_a$ and $n^a_F$ appear in the
combination $\xi_a =
\l_a + n^a_F \l_0$ for all $a$.
The boundary conditions \eq{phibc}\ reflect the
fact that the amplitudes for the gauge field configuration ($\l_b
=\l_0$) at some given value of $n^b_F$ is the same as that of ($\l_b =
0$) at $n^b_F + 1$.

Note that since the differential equation \eq{phieqone}\ is actually
in terms of $\xi_a = \l_a + n^a_F \l_0$ we can write
\be
\phi_{\{ n^a_F \}} (\{ \l_a \}) =  u ( \{ \l_a + n^a_F \})
\label{homthree}
\ee
where $u ( \{ \xi_a \} ) $ is a function of $N$ real variables
$ \xi_a \in (-\infty, \infty)$, satisfying the differential equation
\be
\left [\sum_a -\frac{\del_{\xi_a}^2}{2N^2L} +
H_F^{reg} (\{ \xi_a \}) \right ] u (\{\xi_a \}) = E u( \{\xi_a \} )
\label{ueqone}
\ee
The boundary conditions \eq{phibc} basically ensure the continuity and
smoothness of the function $u(\{ \xi_a \})$ at integer values of
$\xi_a$ (it is enough to ensure continuity of the function and of the
first derivative since the differential equation \eq{ueqone} ensures the
continuity of the higher derivatives). It is indeed the boundary
conditions \eq{phibc}\ that encode the information that $(2\pi \l_a
/\l_0)$'s are `angles' and there is nothing special about the points
$(2\pi \l_a/\l_0) =  0 $ (or $= 2\pi $) to warrant any discontinuity.

Let us now go back to the ansatz for the full wavefunction
\eq{fullstate}\ of the theory. The gauge transformation of the
wavefunction under $\Omega(\{ m_a \})$ can be read off by making the
inverse transformation $\Omega^\dagger$ on the `coordinates' $\psi^a$
and $\l_a$. Using the fact that $ \langle \psi| \Omega | \{n^a_F \}
\rangle = \langle \psi | \{n^a_F - m_a \} \rangle $, we have
\be
\Psi_{n_F}(\psi,\l) \to \Psi_{n_F}^\Omega(\psi,\l)
\equiv \Psi_{n_F}(\psi^{\Omega^\dagger},
\l^{\Omega^\dagger}_a )
=  | \{ n^a_F - m_a \}  \rangle
\otimes \phi_{ \{ n^a_F \} } ( \{\l_a - m_a \} )
\label{gaugetrone}
\ee
Using the boundary conditions \eq{phibc}\
we can rewrite \eq{gaugetrone}\ as
\be
\Psi \to \Psi^\Omega(\psi,\l) \equiv
\Psi(\psi^{\Omega^\dagger}, \l^{\Omega^\dagger})
= | \{ n^a_F - m_a \} \rangle \otimes \phi_{ \{ n^a_F - m_a\} }
(\{ \l_a \})
\ee
The last equation not only tells us that
the state \eq{fullstate}\ is not
gauge-invariant but it also immediately tells us what the gauge
invariant state is. It is simply
\be
|\Psi \rangle =
\sum_{ \{n^a_F \} } | \{ n^a_F \} \rangle \otimes
\phi_{\{n^a_F \}}( \{\l_a \})
\label{fullstatea}
\ee
To see the implication of  the above,
let us calculate the expectation value of the
full hamiltonian $H$ in the state $| \Psi \rangle$.
\be
\ba
\langle \Psi | H | \Psi \rangle
= \sum_{ \{ n^a_F \} } \prod_a \int_0^{\l_0} d\l_a
\phi_{\{n^a_F \}}^*(\{\l_a \})H_{\rm eff}
\phi_{\{n^a_F \}}(\{ \l_a \}) \\
\;\;~~~~~~ = \prod_a \int_{-\infty}^\infty d\xi_a u^*(\{ \xi_a \})
\left( \sum_a -{1\over 2N^2L}{\del^2 /
\del \xi_a^2} + H^{reg}_F ( \{ \xi_a \}
\right) u(\{ \xi_a \})
\ea
\label{decompact}
\ee
where $H_{\rm eff}$ is given by the expression in square brackets on
the left hand side of \eq{phieqone}.  The last line shows that the
presence of fermions forces a {\em decompactification} of the
eigenvalues $\l_a$ to the gauge-invariant combination $\l_a + n^a_F =
\xi_a$ which is a real number $\in (-\infty, \infty)$. It also tells
us that there is no fixed fermi level in a compact gauge theory with
dynamical gauge fields; rather, the effective theory of the gauge
fields $\{\l_a\}$ in the fermi vacuum is given by a density matrix
$\rho(n^a_F, \l_a | n^{\prime a}_F, \l'_a)$.  The trace wih respect to
the density matrices in the above calculations reduces to a single sum
over $ \{ n^a_F \} $ because quantities like total energy etc. in the
large $N$ limit are diagonal in the occupation number representation.

\vspace{2 ex}

\noindent\underline{Connection with axial anomaly:}\hfill\break
\noindent
The above discussion has an intimate connection with the existence
of the well-known nonabelian axial anomaly in the present theory
\cite{BARDEEN}
\be
D^\mu J_{\mu,5} =  { g \over 2\pi } \epsilon^{\mu\nu} F_{\mu\nu}
\label{anomone}
\ee
where $J_{\mu, 5}^{ab} = {\bar \psi}^a \gamma^\mu \gamma^5 \psi^b$ is
the axial colour current. Let us denote by $Q^a_5$ the diagonal axial
charges $\int_0^L J^{aa}_{0,5} = \int_0^L \psi^{\dagger a} \gamma^5
\psi^a$. Integrating both sides of \eq{anomone}\ over all space and
restricting to diagonal colour components we get (in the gauge
$\del_x A_1 =0,\; A_1^{ab} = \l_a \delta_{ab} $)
\be
\del_t Q^a_5  + ig \int_0^L dx\, [ A_0, J_{0,5} ]^{aa} =
   { g \over \pi } \int_0^L dx\, F_{01}^{aa}
\label{anomtwo}
\ee
The commutator term on the left hand side vanishes in the fermi
vacuum, while the integral on the right hand side is just the
constant colour-diagonal component of the transverse electric
field. With these inputs we may rewrite \eq{anomtwo}\  as
\be
\del_t Q^a_5 = \frac{2 E_a}{\l_0 }
\label{anomthree}
\ee
where we have denoted $(1/L) \int_0^L dx\, F_{01}^{aa}$ by $E_a$.

Let us now see how \eq{anomthree}\ arises from our discussion
of the previous section. In the fermi vacuum, we have
\be
\ba
Q^a_5 = Q^a_R - Q^a_L,\\
Q^a_R =  \sum_{n = -\infty}^{n^a_F} 1, \quad
Q^a_L =  \sum_{n = n^a_F +1 }^{\infty} 1
\ea
\ee
Using the exponential regularization once again, we get
\be
Q^a_R(\epsilon) = {1\over \epsilon} + {1 \over 2} + n^a_F +
{\l_a \over \l_0} + o(\epsilon)
= {1 \over \epsilon} + {1 \over 2} + { \xi_a \over \l_0}
+ o(\epsilon)
\ee
and
\be
Q^a_L(\epsilon) = {1\over \epsilon} - {1 \over 2} - n^a_F
- {\l_a \over \l_0} + o(\epsilon)
= {1 \over \epsilon} - {1 \over 2} - { \xi_a \over \l_0}
+ o(\epsilon)
\ee
giving
\be
Q^a_5 =  {2 \xi_a \over \l_0}  + 1
\label{anomfour}
\ee
Thus we see that the regularized diagonal axial charges, $Q^a_5$, depend
on the gauge fields $\l_a$. In fact, as $\l_a$ is tuned from $0$ to
$\l_0$ (eqs. \eq{homotopy} and \eq{homtwo}
), $Q^a_5$ changes by 2.  The reason for this is the
spectral flow which causes an existing left-handed energy level to be
shifted to the right-handed energy levels. As a consequence of this
spectral flow, the regularized diagonal axial charges are neither
integers nor conserved. The best way to see the presence of anomaly
(that is, nonconservation of $Q^a_5$) is
to work in terms of the extended variable $\xi_a$ (see the
example in \eq{decompact})
and to work out the Heisenberg operator equation of motion for the
charges $Q^a_5$ in the effective theory of the gauge fields.  We get
\be
{i \over N} \del_t  Q_5^a = [Q^5_a,  - \sum_a \frac{ \del_{\xi_a}^2}
{ 2N^2 L} + H_F^{reg} ]
\label{anomfive}
\ee
Using \eq{anomfour} and \eq{anomfive} we get
\be
\del_t Q^5_a =  - i( \bar g/\pi N) \del_{\xi_a} =   2 E_a / \l_0
\label{anomsix}
\ee
which is identical to \eq{anomthree}.\footnote{In the last equality
of \eq{anomsix} we have identified $E_a$ with the
operator $(-i/N) \del_{\xi_a}$. This is easy to see because
on the wavefunctions (eq.
\eq{homthree}) the operator $\del_{\l_a}$ reduces
to $\del_{\xi_a}$ in the sense of eq. \eq{decompact}.}

\vspace{5 ex}

\noindent {\large\bf 6. The effective theory
of gauge degrees of freedom ---  description in
terms of fermion bilocal operator}

\vspace{5 ex}

\def\l{\xi}
We have seen that the effective theory of gauge degrees of
freedom is described
by a system of $N$ nonrelativistic fermions%
\footnote{The mapping of the 1-plaquette lattice gauge theory into
free non-relativistic fermions is an old result \cite{SRW}. The method
of proof used in \cite{SRW} can be trivially extended to any polygon
approximation of the circle. In the continuum limit, as expected, the
hamiltonian is given by the laplacian on the unitary group. The
wave-functions are $U(N)$ singlets in the absence of external
sources. A recent discussion of YM theory on circle is given in
\cite{MINAHAN}.}
interacting with each other. The effective hamiltonian is given by
\be
\ba
H_{\rm eff}
=  - \sum_a {1 \over 2 L N^2 } \del_{ \xi_a}^2 +
{\bar g \lambda_0 \over N }\sum_a ( \xi_a/\lambda_0 + 1/2)^2
\\
\,~~~~~~~~~~~~~~~ - {1 \over 2 L \lambda_0^2 N^2} \sum_{a,b}
\left[ \psi(w_{ab}) + \psi( - w_{ab}) + w_{ab} \{ \psi'(w_{ab})
-   \psi'(- w_{ab}) \} \right], \quad w_{ab} =
(\xi_a - \xi_b)/\lambda_0
\ea
\label{heffa}
\ee
Introducing a new fermi field, $\eta (\xi, t)$, which
satisfies the anticommutation relation
\be
\{ \eta (\l, t), \eta^\dagger (\l', t) \} = \delta (\l - \l'),
\label{fourfiveone}
\ee
we can build a field theoretic description for the effective theory:
\be
{\cal H} = \int^{\infty}_{-\infty} d\l\; \eta^\dagger (\l, t)~h_\xi
\eta (\l, t) - \int_{-\infty}^\infty d\xi\,\int_{-\infty}^\infty
d\xi'\, \eta^\dagger(\xi)\eta(\xi) K_{\xi\xi'}
\eta^\dagger(\xi')\eta(\xi')
\label{fourfivetwo}
\ee
where
\be
h_\xi = -{ 1\over 2 L N^2 } \del^2_\xi + {\bar g \lambda_0 \over N}
\big( { \xi \over \lambda_0} + { 1 \over 2 }\big)^2
\label{calhdef}
\ee
and
\be
K_{\xi\xi'} = K( \xi - \xi'), \quad
K(\xi) = {1 \over 2L\lambda_0^2 N^2}(1 + \xi\del_\xi)(\psi ( \xi )
+ \psi( - \xi))
\label{calhdefa}
\ee
The physical states of this system are required to satisfy the
operator constraint
\be
\int^{\infty}_{-\infty}
d\l \; \eta^\dagger (\l, t) \; \eta(\l, t) = N
\label{fourfivethree}
\ee
which imposes the requirement of a fixed number ($N$) of fermions.

Because of the constraint \eq{fourfivethree}\ the physical states of
this system are bosonic.  A procedure for complete bosonization of
systems like the present one has been developed in
\cite{COADJOINT,PATHINTEGRAL}.  The key ingredient in terms of which
the bosonization is achieved is the bilocal operator $\Phi (t)$,
defined by
\be
\hat\Phi_{\l\l'}(t) = \eta(\l,t)\; \eta^\dagger (\l',t).
\label{fourfivefour}
\ee
The bilocal operator satisfies a $W_\infty$ algebra, which can be
deduced from the fermion anticommutation relation \eq{fourfiveone}:
\be
\left[\hat \Phi_{\l_1\l_2}(t), \; \hat\Phi_{\l'_1\l'_2}(t) \right] =
\delta(\l_2 - \l'_1) \; \hat\Phi_{\l_1\l'_2}(t) - \delta(\l_1 -
\l'_2) \; \hat\Phi_{\l'_1\l_2} (t)
\label{fourfivefive}
\ee
It also satisfies a quadratic constraint, which follows directly from
\eq{fourfivethree}:
\be
\hat\Phi^2_{\l\l'} (t) \equiv \int^{\infty}_{-\infty} d\l''\;
\hat\Phi_{\l\l''}(t) \; \hat\Phi_{\l''\l'} (t) = (N+1) \;
\hat\Phi_{\l\l'} (t), \;\; {\rm i.e.} \; \hat\Phi^2 = (N+1) \;
\hat\Phi.
\label{fourfivesix}
\ee
Also in terms of the bilocal operator $\hat\Phi$ the constraint
\eq{fourfivethree}\ reads
\be
Tr (1 - \hat\Phi) = N
\label{fourfiveseven}
\ee
where the symbol `$Tr$' stands for the usual matrix trace, i.e. $TrA =
\int^{\infty}_{-\infty} d\l A_{\l\l}$.

A description of the classical phase space of this system is obtained
in terms of the expectation values $\langle \hat\Phi \rangle$ of the
bilocal operator in $W_\infty$-coherent states.  We shall denote these
by $\Phi$.  One can show that the classical phase space satisfies the
constraints
\be
\Phi^2 = \Phi, \;\; Tr (1 - \Phi) = N.
\label{fourfiveeight}
\ee
These constraints are classical analogues of
the operator constraints \eq{fourfivesix}\ and \eq{fourfiveseven} and
define a coadjoint orbit of the $W_\infty$ algebra \eq{fourfivefive}.
One can now immediately write down a bosonic field theory action,
following Kirillov's method of coadjoint orbits:
\be
S_{\rm eff} = {i\over N}
\int_\Sigma dsdt \; Tr \left(\Phi (s,t) \left[\partial_t \Phi(s,t),
\partial_s \Phi (s,t)\right]\right) -
\int dt \; Tr(h \Phi) + \int dt \int_{-\infty}^\infty
d\xi \int_{-\infty}^\infty d\xi'
\Phi_{\xi'\xi} \Phi_{\xi\xi'} K_{\xi\xi'}
\label{fourfivenine}
\ee
where
\be
h_{\xi\xi'} =  h_\xi \delta(\xi - \xi')
\label{fourfiveten}
\ee
The integration region $\Sigma$ of $(s,t)$ is the lower half plane,
$s\epsilon (-\infty, 0), t\epsilon (-\infty, +\infty)$, and $\Phi
(s,t)$ satisfies the boundary conditions $\Phi (s = 0, t) = \Phi
(t),\; \Phi(s,t) \Big|_{s \rightarrow -\infty} =$
time-independent constant matrix. We note that the
term quadratic in $\Phi$ in the action  \eq{fourfivenine}\ breaks
the residual part of $W_\infty$ invariance that survives after
introduction of the `magnetic field' $h$.

Because of the constraints \eq{fourfiveeight}, the allowed variations in
$\Phi$ are the $W_\infty$ rotations:
\be
\Phi \rightarrow e^{iW} \Phi \; e^{-iW} \approx \Phi + i [W, \Phi] +
\ldots
\label{fourfiveeleven}
\ee
Making such a variation in $S_{\rm eff}$ we deduce the classical
equation of motion for $\Phi$:
\be
{i\over N}
\partial_t \Phi_{\xi\xi'}(t) = \left[\Phi (t), h\right]_{\xi\xi'}
+  \int_{-\infty}^\infty d\xi'' \;
\Phi_{\xi\xi''} \Phi_{\xi''\xi'} (K_{\xi\xi''} - K_{\xi''\xi'})
\label{fourfivetwelve}
\ee
One can easily check that this is identical to the operator equation
obtained using \eq{fourfivetwo}. Note that the factor of $1/N$
on the left hand side is there because $\hbar = 1/N$.

The classical ground state of this system is described by  such a
time-independent solution of \eq{fourfivetwelve} that satisfies the
constraints \eq{fourfiveeight}.  Excited states are described by
arbitrary $W_\infty$ rotations of the ground state.  In general the
description of these states is rather complicated.  However, the
semiclassical limit of a theory of the type described by the action
\eq{fourfivenine} and constraints \eq{fourfiveeight} admits of
a simple fermi fluid description in the sense of a Hartree-Fock
approximation.  If we further restrict ourselves to small fluctuations
of the fermi surface of a special type (`quadratic profile') then the
fermi fluid description \cite{DMWCLASS}\ further simplifies to a
description in terms of the collective variables \cite{JEVSAK}, the
density $\rho(\l,t)$ and its conjugate momentum $\Pi(\l, t)$.  The
semiclassical limit of the present theory is obtained when $N \to
\infty$.

In the small $L$ limit one can neglect the quartic fermion term in
\eq{fourfivetwo}\ and the theory considerably simplifies.
We shall discuss this limit separately in a later section.

\vspace{5 ex}

\noindent {\large\bf 7. The operator algebra of meson fluctuations}

\vspace{5 ex}

\def\l{\lambda}
In the previous sections we fixed our attention on the lowest energy
state in the fermionic sector since fermionic fluctuations do not
affect the effective theory of gauge fields in leading order in $N$.
To obtain the meson spectrum in this theory, however, we need to
consider fluctuations in the bilocal field. The allowed fluctuations
are $\widetilde {W}^f_\infty$-rotations. Let us denote  the
expectation value of $\M(\l)$ in the full vacuum of the theory by
$\M_0(\l)$ and the corresponding quantity for ${\cal M}(\l)$
(see eqn. \eq{calmdef}) by ${\cal M}_0(\l)$. Then, for fluctuations
we must write
\be
\ba
{\cal M}(\lambda,t) = e^{
iW(\lambda, t)/\sqrt N} {\cal M}_0(\lambda)
e^{-iW(\lambda, t)/\sqrt N} \\
{}~~~~~~~ = {\cal M}_0(\lambda) +
\frac{i}{\sqrt N}[ W(\l, t), {\cal M}_0(\l) ]
+ {i^2 \over 2! N} [ W(\l, t), [ W(\l, t), {\cal M}_0(\l) ]]
+ o(\frac{1}{N\sqrt N} )
\label{fluctone}
\ea
\ee
Here ${\cal M}(\l)$ is defined as in \eq{calmdef}.
By definition, ${\cal M}_0(\l)$ satisfies the quadratic constraint
${\cal M}^2_0(\l) = {\cal M}_0(\l)$. This implies the existence
of the two projection operators $P_+$ and $P_-$, which are
given by
\be
P_+(\l) = {\cal M}_0(\l), \quad P_-(\l) = 1 - {\cal M}_0(\l),
\quad P_\pm^2 (\l) = P_\pm (\l)
\label{project}
\ee
With the help of these projection operators we can distinguish
four different kinds  of operators that may be constructed
out of ${\cal M}(\l, t)$:
\be
{\cal M}_{\pm\pm} (\l, t) = P_\pm (\l) {\cal M}(\l, t) P_\pm (\l)
\label{mpm}, \quad
{\cal M}_{\pm\mp} (\l, t) = P_\pm (\l) {\cal M}(\l, t) P_\mp (\l)
\label{mmp}
\ee
Using \eq{fluctone}\ and \eq{project}\ we get
\be
{\cal M}_{\pm \mp} (\l, t) = \mp {i \over \sqrt N} W{\pm \mp}(\l,t)
+ o({1\over N})
\label{projc}
\ee
\be
{\cal M}_{++}(\l, t) = {\cal M}_0(\l) - {1 \over N} W_{+-} (\l, t)
W_{-+} (\l,t) + o({1\over N \sqrt N})
\label{projd}
\ee
\be
{\cal M}_{--}(\l, t) = {1 \over N} W_{-+} (\l, t)
W_{+-} (\l,t) + o({1\over N \sqrt N})
\label{proje}
\ee
The $\widetilde {W}^f_\infty$ algebra
satisfied by ${\cal M}(\l)$, eqn. \eq{fourthirtyfive},
implies an algebra for the operators $
{\cal M}_{\pm \pm}(\l), {\cal M}_{\pm \mp}(\l)$. It turns out that the
first two of these generate a subalgebra of $\widetilde {W}^f_\infty$,
which acts on the other two. We give the full algebra below.
\be
\left[{\cal M}^{i\alpha, j\beta}_{\pm \pm xy} (\lambda), \;
{\cal M}^{i'\alpha',
j'\beta'}_{\pm \pm x'y'} (\lambda') \right] =
{\l_0 \over N} \delta_P(\l -
\l')  \left\{
P^{i'\alpha', j\beta}_{\pm x'y} (\lambda)
{\cal M}^{i\alpha, j'\beta'}_{\pm\pm xy'} (\lambda) -
P^{i\alpha, j'\beta'}_{\pm xy'} (\lambda)
{\cal M}^{i'\alpha',j\beta}_{\pm\pm x'y} (\lambda) \right\}
\label{projf}
\ee
\be
\left[{\cal M}^{i\alpha, j\beta}_{\pm \pm xy} (\lambda), \;
{\cal M}^{i'\alpha',
j'\beta'}_{\pm \mp x'y'} (\lambda') \right] =
{\l_0 \over N} \delta_P(\l -
\l') P^{i'\alpha', j\beta}_{\pm x'y} (\lambda)
{\cal M}^{i\alpha, j'\beta'}_{\pm\mp xy'} (\lambda)
\label{projg}
\ee
\be
\left[{\cal M}^{i\alpha, j\beta}_{\pm \pm xy} (\lambda), \;
{\cal M}^{i'\alpha',
j'\beta'}_{\mp \pm x'y'} (\lambda') \right] =
- {\l_0 \over N} \delta_P(\l - \l')
P^{i\alpha, j'\beta'}_{\pm xy'} (\lambda)
{\cal M}^{i'\alpha',j\beta}_{\mp \pm x'y} (\lambda)
\label{projh}
\ee
\be
\left[{\cal M}^{i\alpha, j\beta}_{+ - xy} (\lambda), \;
{\cal M}^{i'\alpha',
j'\beta'}_{- + x'y'} (\lambda') \right] =
{\l_0 \over N} \delta_P(\l -
\l') \left\{
P^{i'\alpha', j\beta}_{- x'y} (\lambda)
{\cal M}^{i\alpha, j'\beta'}_{+ + xy'} (\lambda) -
P^{i\alpha, j'\beta'}_{+  xy'} (\lambda)
{\cal M}^{i'\alpha',j\beta}_{- - x'y} (\lambda) \right\}
\label{proji}
\ee
The commutators of all other combinations vanish. The first
of the above is the statement that ${\cal M}_{\pm \pm}(\l)$
satisfy a subalgebra of $\widetilde {W}^f_\infty$. The
second and third show that this algebra acts on
${\cal M}_{\pm \mp}(\l)$. The last is best understood
in terms of the fluctuations $W_{\pm \mp}(\l)$. Let us
substitute \eq{projc}$-$\eq{proje}\ in the above algebra,
retaining only the first non-trivial term in the fluctuations
in ${1\over \sqrt N}$ expansion. We see that $({\cal M}_{++}(\l)
-  {\cal M}_0(\l))$ and ${\cal M}_{--}(\l)$, which are
realized to this order in ${1 \over \sqrt N}$ in terms of
the fluctuations $W_{+-}(\l)$ and $W_{-+}(\l)$, satisfy a
subalgebra of $\widetilde {W}^f_\infty$ which acts on the
fluctuations. The origin of this action is the commutation
relation in \eq{proji}, which gives for the fluctuations
\be
\left[W^{i\alpha, j\beta}_{+ - xy} (\lambda), \;
W^{i'\alpha', j'\beta'}_{- + x'y'} (\lambda') \right] =
\l_0  \delta_P(\l - \l') P^{i'\alpha', j\beta}_{- x'y} (\lambda)
P^{i\alpha, j'\beta'}_{+ xy'} (\lambda)
\label{projj}
\ee
We see that a $c$-number term has appeared in the algebra of
fluctuations \eq{projj}.
In \cite{DMWQCD}\ a corresponding phenomenon had occurred
for meson fluctuations where the $c$-number term was identified
with the central term of a Heisenberg algebra for the fluctuations.

The $c$-number term in \eq{projj}\ depends on the vacuum expectation
value of ${\cal M}(\l)$. We close this section by outlining briefly
how this can be obtained. First we note that (by equation
\eq{calmdef}) ${\cal M}_0(\l) = \M_0(g^{-1}(\l))$, where $g$ is the
transformation that takes the eigenvalue distribution in the ground
state of the effective gauge theory to the fiducial distribution (see
eqn. \eq{fidu}). Note that the decompactification of the eigenvalues
discussed in section 5 does not affect this since the transformation
$g$ only sees the fractional parts of $\xi_a$'s (or $\l_a$'s). Now,
$g(\l)$ can be expressed in terms of the density $\U(\l)$ as in
eqn. \eq{ffound}. It follows, therefore, that a computation of ${\cal
M}_0(\l)$ involves knowing $\M_0(\l)$ and $\U_0(\l)$, where the latter
is the vacuum expectation value in the effective gauge
theory. Actually, both $\M_0(\l)$ and $\U_0(\l)$ can be obtained from
a knowledge of the vacuum expectation value of a more general density
variable $\rho(\xi, t)$ defined on the entire real line:
\be
\rho(\xi, t) \equiv {1 \over N} \sum_a \delta(\xi - \xi_a(t))
\label{rhodef}
\ee
It is related to $\U(\l, t)$ by
\be
\U(\l, t) = \sum_{m = -\infty}^\infty \rho( \l + m\l_0, t)
\label{rhovsu}
\ee
Also, $\M_{xy}(\l, t)$, given by \eq{modedef}\ and \eq{mvev},
can be rewritten in terms of $\rho(\xi, t)$ as
\be
\M_{xy}(\l, t) = - \gamma^5 {1 \over L \U(\l, t)}\;
{ \widetilde  \rho(\l, t; x-y) \over 1 - e^{-i 2\pi (x -y)/L}}
\label{mvaltwo}
\ee
where
\be
\widetilde \rho(\l, t; x-y) = \sum_{m = -\infty}^\infty
e^{i 2\pi m (x-y)/L} \rho(\l + m\l_0, t)
\label{mvalthree}
\ee
Thus a knowledge of $\rho_0(\xi)$ is sufficient to compute
$\M_0(\l)$ and $\U_0(\l)$ and hence ${\cal M}_0(\l)$. A computation
of the value of $\rho_0(\xi)$ in the small $L$ limit is given
in the next section.

\vspace{5 ex}

\noindent {\large\bf 8. The small $L$ limit}

\vspace{5 ex}

Here we are interested in the limit $L \to 0$.  In this limit $\l_0 =
2\pi/\bar g L \rightarrow \infty$, i.e. as the length of the space
circle becomes vanishingly small, the length of the internal circle
becomes infinitely large. We then expect the dynamics to reside only
in the gauge field degrees of freedom, since the fermionic
fluctuations are further suppressed by factors of $1/L$ in addition to
the $1/N$ suppression already present at large $N$.  Now, the dynamics
of the gauge fields is given by the effective hamiltonian $H_{\rm
eff}$ \eq{heffa}.  The first two terms in $H_{\rm eff}$ are of order
$1/L$.  The last term is subleading in $L$ and does not
contribute. This is because its coefficient vanishes linearly as $L
\to 0$ while the quantity in the square brackets remains finite in
this limit%
\footnote{
This can be seen by using the property of the function $\psi(x)$,
$ \psi(x) = \psi(x+1) - 1/x$.
At small $x$ $\psi(x)$ behaves as $ - 1/x$
since $\psi(1)$ is finite. Using this, it
follows that the quantity in the square brakcets in $H_{\rm eff}$
vanishes as $w_{ab} \to 0$.  It is this last limit that is of
relevance here since the maximum separation between two of the scaled
eigenvalues $\xi_a/\l_0$ and $\xi_b/\l_0$ goes as $\sqrt{L}$.  }.
Therefore to leading order in the $L \to 0$ limit, the effective gauge
field dynamics is described by a set of $N$ free non-relativistic
fermions moving on a line in a harmonic oscillator potential.  This
problem may be solved by the bosonization methods introduced and
extensively discussed in \cite{COADJOINT,PATHINTEGRAL}\ in the closely
related context of the $c=1$ matrix model, in which the only
difference is that the potential has the wrong sign.

In Sec. 6 we have introduced this formalism of bosonization for the
more general problem of interacting fermions. In the small $L$ limit
the quartic fermion term in ${\cal H}$, eqn. \eq{fourfivetwo}, can be
neglected and the theory considerably simplifies.  The relevant
equation of motion in this limit is
\be
{i\over N}\del_t \Phi(t) =  [\Phi(t), h].
\label{eqforphi}
\ee
The classical ground state of this system, $\Phi_0$, is described by a
time-independent solution of \eq{eqforphi}\ which satisfies the
constraints
\be
\Phi_0^2 = \Phi_0, \quad Tr (1 - \Phi_0) = N
\label{something}
\ee
Let us parametrize such a solution as
\be
\Phi_{0,\xi\xi'} = \delta(\xi - \xi') - \sum_{n=0}^\infty c_n \phi_n(\xi)
\phi^*_n(\xi')
\label{fourfivethirteen}
\ee
This  satisfies the equation of motion \eq{eqforphi} provided
the $\phi_n$'s satisfy the Schrodinger equation for a harmonic
oscillator:
\be
\left[ - {1 \over 2 L N^2} \del_{\xi}^2
+ { \bar g \l_0 \over N} ( { \xi \over \l_0 } + { 1 \over 2})^2
\right] \phi_n = E_n \phi_n
\label{fourfivefourteen}
\ee
Assuming that $\phi_n$'s are orthonormal and imposing the constraints
\eq{something}\ on \eq{fourfivethirteen}, we get
\be
c^2_n = c_n, \; \sum_{n=0}^\infty c_n = N
\label{fourfivefifteen}
\ee

The first of \eq{fourfivefifteen}\ says that any given level is
either occupied or unoccupied (fermions!), while the second says that
precisely $N$ levels are occupied.  Since the energy of the level
labelled by $n$ goes as $n$, minimising energy in the ground state
implies that the first $N$ levels are occupied, {\it i.e.},
\be
c_n = \theta( -n + N -1).
\label{cvev}
\ee
The density
$\rho(\xi,t)$ is given by the formula
\be
\rho(\xi,t) = {1 \over N}\left(1 -
\Phi(t)\right)_{\xi\xi}.
\ee
For the ground state we see that
\be
\rho_0(\xi) = \frac{1}{N}
\sum_{n=0}^{N-1} \phi_n(\xi)\phi^*_n(\xi).
\ee

Excited states (glueballs) are described by $W_\infty$ rotations
of the ground state constructed above. Writing
\be
\Phi(t) = e^{i\Delta(t)} \Phi_0 e^{-i \Delta(t)}
\label{phifa}
\ee
and neglecting quadratic and higher terms in $\Delta(t)$, we
see that $\Delta(t)$ satisfies the equation of motion
\be
[ {i\over N}\del_t \Delta(t) - [ \Delta(t), h ], \Phi_0 ] = 0
\label{phifb}
\ee
This leads to
\be
{i\over N}\del_t \Delta_{\pm \mp}(t) =  [ \Delta_{\pm \mp} (t), h]
\label{phifc}
\ee
where
\be
\Delta_{+-} (t) \equiv \Phi_0  \Delta(t) (1 - \Phi_0), \quad
\Delta_{-+} (t) \equiv (1 - \Phi_0 ) \Delta(t)\Phi_0
\label{phifd}
\ee
This construction is identical to the one just described for meson
fluctuations and everything discussed there also applies
here. Defining
\be
\Delta_{\pm\mp}^{nm}(t) \equiv \int_{-\infty}^\infty d\xi
\int_{-\infty}^\infty d\xi' \;
\phi^*_n(\xi) (\Delta_{\pm \mp}(t))_{\xi\xi'}
\phi_m(\xi')
\label{phife}
\ee
and using the ground state solution $\Phi_0$ obtained above one can
show that $\Delta_{+-}^{nm}(t)$ vanishes if $n< N$ or $m \ge N$.
Similarly, one can also show that $\Delta_{-+}^{nm}(t)$ vanishes
if $n \ge N$ or $m < N$. Moreover, these two satisfy the algebra
\be
[ \Delta_{+-}^{nm},  \Delta_{-+}^{n'm'} ]= \theta(n-N)\theta(-m+N-1)
\delta^{nm'}\delta^{n'm}
\label{phiff}
\ee
Thus, for appropriate ranges of indices, $\Delta_{+-}^{nm}$
and $\Delta_{-+}^{mn}$ form a canonically conjugate pair.

The equation of motion for $\Delta_{\pm \mp}^{nm}$ is
\be
{i \over N} \del_t \Delta_{\pm \mp}^{nm}(t) =
- (E_n - E_m) \Delta_{\pm \mp}^{nm}(t)
\label{phifg}
\ee
which is trivially solved by
\be
\Delta_{\pm \mp}^{nm}(t)
= e^{iN (E_n - E_m)t} \Delta_{\pm \mp}^{nm}(0)
\label{phifh}
\ee
Thus the spectrum of glueballs consists of `discrete' states labelled
by two integers $(n,m)$ \cite{DISCRETE}. A residual wedge subalgebra
of the original $W_\infty$ acts on this tower of states and moves them
around.  This subalgebra is generated by
\be
\Phi_{++} \equiv \Phi_0 \Phi \Phi_0,
\label{phifi}
\ee
and
\be
\Phi_{--} \equiv (1 - \Phi_0) \Phi (1 - \Phi_0),
\label{phifj}
\ee
exactly like the situation discussed in the previous section.
$\Phi_{++}^{nm}$ is nonvanishing only for $n,m \ge N$, while
$\Phi_{--}^{nm}$ is so for $n,m < N$.
Their action on $\Delta_{+-}$ is given by
\be
[ \Phi_{++}^{nm}, \Delta_{+-}^{n'm'} ]
= \theta(n' - N) \delta^{n'm} \Delta_{+-}^{nm'}
\label{phifk}
\ee
\be
[ \Phi_{--}^{nm}, \Delta_{+-}^{n'm'} ]
= -\theta(-m' +N-1) \delta^{nm'} \Delta_{+-}^{n'm}
\label{phifl}
\ee
There is a similar action on $\Delta_{-+}^{nm}$. Thus the spectrum
in the limit $L \to 0$ consists of a tower of discrete states
(glueballs) which realize a representation of a wedge subalgebra
of $W_\infty$ algebra.

\vspace{5 ex}

\noindent {\large\bf 9. The large $L$ limit}

\vspace{5 ex}

In the limit $L \to \infty$ the pure Yang-Mills part of the
hamiltonian behaves like a system of infinitely massive particles.
Fluctuations in $\{ \l_a \}$ are therefore small, and so we can fix
our attention on the ground state only in the Yang-Mills sector.
Also as $L \to \infty$,
$\l_0 = 2\pi/\bar gL \rightarrow 0$, i.e. the length of
internal circle becomes vanishingly small.  We then expect only the
zero mode of $\M(\l)$ to survive.

Actually, in this limit $\M(\l)$ is not a convenient variable to work
with since it is not periodic in $\l$.  However, $\widetilde M(\l)$,
to which $\M(\l)$ is related as in \eq{fourten}, is periodic and so
this is the appropriate variable to work with in the large $L$ limit.
In terms of $\widetilde M(\l)$ the equation of motion
\eq{fourthirtyeight}\ reads
\be
\ba
i\left(\partial_t - (1/L)\sum_a p_a \del_{\l_a}
\right) \; \widetilde M^{i\alpha, j\beta}_{xy} (\l, t) \\
{}~~~=
\left[-\gamma^5 i\partial + m\gamma^0, \widetilde M(\l,
t)\right]^{i\alpha, j\beta}_{xy} + {\bar g^2 \over 2} \int^{\l_0}_0
d\l'\;\; \U(\l',t) \int^L_0 dz \Big[\widetilde K(\l'-\l; x-z) \\
{}~~~~~\left(\widetilde
M_{xz}(\l',t) \widetilde M_{zy}(\l,t)\right)^{i\alpha,j\beta}
- \widetilde K(\l-\l'{y-z}) \left(\widetilde M_{xz}(\l, t) \widetilde
M_{zy}(\l',t)\right)^{i\alpha,j\beta}\Big],
\label{fourfortytwo}
\ea
\ee
where $\widetilde K(\l;x) = e^{-i\bar gx\l} \bar K(\l;x), \bar
K(\l;x)$ being given by \eq{fourfortyone}.  Moreover, $\widetilde
M(\l)$ satisfies constraints that are identical to \eq{fourfourteen}\
and \eq{fourfifteen}:
\be
\widetilde M^2(\l) = \widetilde M(\l),
\label{fourfortyfour}
\ee
\be
Tr(1 - \widetilde M(\l)) = 0.
\label{fourfortyfive}
\ee

Let us now expand $\widetilde M(\l,t)$ in modes
in $\l$:
\be
\widetilde M(\l,t) = \sum^{+\infty}_{n=-\infty} e^{i{2\pi n\l \over \l_0}}
\widetilde M^{(n)}(t).
\label{fourfortyeight}
\ee
Retaining only its zero mode in the equation of motion
\eq{fourfortytwo}\  and in the
constraints \eq{fourfortyfour}\ and \eq{fourfortyfive}, we get
\be
i\partial_t\widetilde M^{(0)}(t) = \left[ - \gamma^5 i\partial + m\gamma^0,
\widetilde M^{(0)} (t) \right] + {\bar g^2 \over 2} \left[\widetilde
M^{(0)}(t), \widetilde M'^{(0)}(t)\right],
\label{fourfifty}
\ee
\be
\left(\widetilde M^{(0)}\right)^2 = \widetilde M^{(0)},
\label{fourfiftyone}
\ee
\be
Tr (1 - \widetilde M^{(0)}) = 0,
\label{fourfiftytwo}
\ee
where $\widetilde M'^{(0)}_{xy} (t) = \widetilde M^{(0)}_{xy} (t)
|x-y|$.  These are precisely the equations of motion and the
constraints that one gets in the gauge $A_1 = 0$ for the fermion
bilocal for QCD$_2$ on a plane.  The nonzero modes of $\widetilde
M(\l)$ decouple because excitation of these modes makes the
corresponding states infinitely heavy in the limit $L\rightarrow
\infty$.

We mention here that in \cite{DMWQCD} we solved the model defined by
the equations \eq{fourfifty}, \eq{fourfiftyone} and \eq{fourfiftytwo}
in the manifestly Lorentz-invariant gauge $A_+ =0$. We showed there
that the spectrum of excitations of this model is identical to the
'tHooft spectrum for mesons and that these particles realize a
representation of a wedge subalgebra of the $W_\infty$-algebra which
the fermion bilocal fields satisfy.

\vspace{5 ex}

\noindent {\large\bf 10. Concluding Remarks}

\vspace{5 ex}

To conclude we remark that some of the ideas presented in this paper
may well apply to finite temperature QCD. It is well-known that field
theories at finite temperature ($T= 1/ \beta$) can be mapped to
Euclidean field theories with compactified time (with period
$\beta$). On the other hand, for Euclidean field theories, time is
just any one of the dimensions.  Therefore, if we consider the
Euclidean version of our theory, then we can learn about the spectrum
of glueballs and mesons at $\beta \to 0$ and $\beta \to \infty$ from
our analysis at $L \to 0$ and $L \to \infty$ respectively. Indeed this
is perhaps the simplest model involving dynamical quarks and gauge
fields where one can learn about their excitations and interactions at
finite temperature in an analytic fashion. Implications of this
model for QCD at finite temperature are being worked out.

The second remark is that by tuning $L$ in this theory we seem to
achieve a `melting' of topological degrees of freedom to propagating
degrees of freedom. As we have seen, in the small $L$ limit ($\beta
\to 0$) mesons are frozen, and the theory is described in terms of the
quantum mechanical degrees of freedom ($\l_a(t)$) which do not depend
on space. As we keep increasing $L$, the mesonic fluctuations come
into play. It is a very interesting open question how the theory
behaves at intermediate values of $L$.

The third remark is about what we can learn from this model {\it
vis-a-vis} string theory. The interaction between glueballs and mesons
in this theory can be compared with an interaction between closed and
open strings respectively where the closed string sector depends only
on time and not on space.  The freezing of the propagating (open
string) degrees of freedom at $L \to 0$ tempts one to identify the
remaining, purely quantum mechanical, closed string degrees of freedom
as the degrees of freedom of a `topological' string theory. There is
also the interesting possibility that  this theory has a three
dimensional $(x, t, \xi)$ description where $\xi$ is a new dimension
arising out of the eigenvalues in the gauge sector. We leave these
interesting issues for future work.

\end{document}